\documentclass[11pt, reqno]{amsart}
\usepackage{amsfonts,latexsym}
\usepackage{amsmath}
\usepackage{amscd}
\usepackage{float,amsmath,amssymb,mathrsfs,bm,multirow,graphics}
\usepackage[dvips]{graphicx}
\usepackage[percent]{overpic}

\newcommand{\nequation}{\setcounter{equation}{0}}
\renewcommand{\theequation}{\mbox{\arabic{section}.\arabic{equation}}}
\newcommand{\R}{{\Bbb R}}

\newcommand{\C}{{\Bbb C}}
\newcommand{\Z}{{\Bbb Z}}

\newcommand{\proofbegin}{\noindent{\it Proof.\quad}}
\newcommand{\proofend}{\hfill$\Box$\bigskip}

\newcommand{\res}{\text{\upshape Res\,}}

\newtheorem{theorem}{Theorem}[section]
\newtheorem{proposition}[theorem]{Proposition}
\newtheorem{lemma}[theorem]{Lemma}

\newtheorem{assumption}[theorem]{Assumption}
\newtheorem{remark}[theorem]{Remark}

\newtheorem{figuretext}{Figure}

\input epsf
\title[Degasperis-Procesi equation on the half-line]
{The Degasperis-Procesi equation on the half-line}

\author{Jonatan Lenells}
\begin{document}
\maketitle

\vspace{-.6cm}
\begin{center} \small
Department of Mathematics, Baylor University, \\
One Bear Place \#97328, Waco, TX 76798, USA. \\
E-mail: Jonatan\_Lenells@baylor.edu \\
Phone: +1 254 710 1103
\end{center}

\begin{abstract} 
\noindent
We analyze a class of initial-boundary value problems for the Degasperis-Procesi equation on the half-line. Assuming that the solution $u(x,t)$ exists, we show that it can be recovered from its initial and boundary values via the solution of a Riemann-Hilbert problem formulated in the plane of the complex spectral parameter $k$. 
\end{abstract}

\bigskip



%\tableofcontents

\section{Introduction}\nequation
The Degasperis-Procesi (DP) equation
\begin{equation}\label{DP}
  u_t - u_{txx} + 3\kappa u_x+4uu_x - 3u_xu_{xx} - uu_{xxx}=0,
\end{equation}
where $u(x,t)$ is a real-valued function and $\kappa > 0$ is a parameter, was found in \cite{DP1999} using methods of asymptotic integrability. Equation (\ref{DP}) is similar in form to the Camassa-Holm (CH) equation \cite{CH1993} and arises, just like CH, under certain circumstances as a model for water waves propagating over a flat bed \cite{CL2009, J2002}. A Lax pair and a bi-Hamiltonian structure for (\ref{DP}) were presented in \cite{DHH2002}, where the existence of peakon solutions was also established.
An interesting aspect of (\ref{DP}) is the existence of weak solutions with a very low degree of regularity \cite{CK2006}. In particular, along with the peakons, equation (\ref{DP}) also admits `shock-peakons' \cite{Lu2007}. These are discontinuous generalizations of the peakons, which form when a peakon collides with an antipeakon.

Despite the many similarities between DP and CH, the spectral analysis of the corresponding Lax pairs is quite different due to the fact that the isospectral problem of CH is a second-order ODE, whereas that of DP is a third-order ODE. Thus, although the application of the inverse scattering transform (IST) to CH has been studied extensively (see \cite{CM1999} for the periodic case; \cite{BSS1998, BS2006, C2001, CL2003} for the case on the line; and \cite{BS2008} for the case on the half-line), the implementation of the IST to DP has proved to be more intricate. Nevertheless, an inverse scattering approach for computing the $n$-peakon solutions of DP was presented in \cite{LS2003}, and it was recently shown in \cite{CIL2010} (see also \cite{BS2011}) that the solution of the Cauchy problem on the line with initial data $u_0(x) = u(x, 0)$ satisfying $u_0 - u_{0xx} + \kappa > 0$ can be given in terms of the solution of a Riemann-Hilbert (RH) problem, whose jump matrix is specified by $u_0(x)$.

An important recent development in soliton theory has been the generalization of the IST formalism from initial value to initial-boundary value problems developed by Fokas and his collaborators \cite{F1997, F2002} (see also \cite{Fbook}). Initial-boundary value problems appear in many applications, where it is often more natural to assume that the space variable is defined only on part of the real axis. 
In this paper, we consider a class of initial-boundary value problem for equation (\ref{DP}) on the half-line, that is, in the domain
\begin{equation}\label{halflinedomain}
   \Omega = \left\{(x,t) \in \R^2\, |\, 0 \leq x < \infty, \; 0 \leq t< T\right\},
\end{equation}
where $T < \infty$ is a given positive constant. Assuming that a solution exists, we show that $u(x,t)$ can be recovered from the initial and boundary values $u_0(x), g_0(t), g_1(t)$, $g_2(t)$ defined by
\begin{align}\nonumber
& u_0(x) = u(x, 0), \quad 0 < x < \infty, 
	\\ \label{boundaryvalues}
& g_0(t) = u(0,t), \quad g_1(t) = u_x(0,t),  \quad g_2(t) = u_{xx}(0,t),\quad 0 < t < T.
\end{align}	
The main peculiarity compared with other applications of the approach of \cite{F1997, F2002} is that the Lax pair involves $3 \times 3$-matrices instead of $2 \times 2$-matrices. This difference leads to some new challenges. We will overcome these challenges by employing an extension of the approach of \cite{F1997, F2002} which was recently developed and implemented to an integrable model PDE with a $3 \times 3$ Lax pair in \cite{L2012}. 

Apart from the $3 \times 3$ Lax pair, the spectral analysis of equation (\ref{DP}) on the half-line also presents some other peculiarities: (a) The presence of singularities in the Lax pair implies that it is necessary to introduce two sets of eigenfunctions. The eigenfunctions in the first set are well-behaved near the points $K_j = e^{\frac{\pi i j}{3} - \frac{\pi i}{6}}$, $j = 1, \dots, 6$, on the unit circle, but have singularities at $k = \infty$ and $k = 0$. The eigenfunctions in the second set are well-behaved near $k = \infty$ and $k = 0$, but have singularities at the $K_j$'s. Together these two sets of eigenfunctions can be used to formulate a RH problem. An analogous situation occurs in the analysis of CH on the half-line in \cite{BS2008} where two sets of eigenfunctions are also used.
(b) The definition of the above eigenfunctions requires certain transformations of the Lax pair that involve a matrix $P(k)$ whose inverse is singular at the sixth roots of unity $\varkappa_j = e^{\frac{\pi i (j-1)}{3}}$, $j = 1, \dots, 6$. Consequently, the basic matrix eigenfunctions which are natural candidates for the formulation of a RH problem, are singular near these points. Following \cite{BS2011}, we overcome this problem by formulating an associated vector RH problem, for which these singularities are absent. 
(c) The formulation of the RH problem depends, in addition to the variables $(x,t)$, on a function $y(x,t)$ which is unknown from the point of view of the inverse problem. In order to obtain a RH problem whose jump matrix involves only known quantities, we have to reparametrize the $x$ variable. This implies that we only obtain a parametric representation for the solution $u(x,t)$.
This type of reparametrization occurs also in the analysis of CH \cite{BS2006, CL2003}, DP \cite{CIL2010, BS2011}, and the generalized sine-Gordon equation studied in  \cite{LFgsg}.

We will consider the class of initial-boundary value problems for (\ref{DP}) for which the initial and boundary values satisfy
\begin{subequations}\label{qassumptions}
\begin{align} 
 & u_0(x) - u_{0xx}(x) + \kappa > 0, \qquad x \geq 0, 
	\\
&  g_0(t) - g_2(t) + \kappa > 0, \qquad 0 \leq t < T,
\end{align}
\end{subequations}
as well as
\begin{align}\label{g0assumption}
  g_0(t) \leq 0,  \qquad 0 \leq t < T.
\end{align}
It is shown in Appendix \ref{Aapp} that the assumptions in (\ref{qassumptions}) imply the following positivity condition which is needed for the spectral analysis:
\begin{align}\label{qpositive}
  u(x,t) - u_{xx}(x,t) + \kappa > 0, \qquad 0 \leq x < \infty, \quad 0 \leq t< T.
\end{align}
The necessity of the condition (\ref{qpositive}) is not related to the half-line domain---an analogous condition is required also for the spectral analysis of DP and CH on the line cf. \cite{BS2008, BS2011, C2001, CIL2010}.
The assumption (\ref{g0assumption}) is used to ensure boundedness of certain eigenfunctions (see Proposition \ref{tildePhiprop} below).

Let us finally point out that in the case of vanishing Dirichlet boundary conditions, local well-posedness for (\ref{DP}) on the half-line was established in \cite{EY2007} by considering an odd extension to the real line.

In Section \ref{laxsec}, we introduce a Lax pair for equation (\ref{DP}) and transform it appropriately. 
In Section \ref{eigensec}, we define eigenfunctions which can be used for the formulation of a RH problem.
In Section \ref{specsec}, we derive expressions for the jump matrices in terms of suitable spectral functions. 
In Section \ref{residuesec}, we derive residue conditions for the pole singularities of the eigenfunctions.
In Section \ref{RHsec}, we state our main result: Under the assumptions (\ref{qassumptions}) and (\ref{g0assumption}), the solution $u(x,t)$ of (\ref{DP}) on the half-line can be reconstructed from the initial and boundary data via the solution of a RH problem. 

\section{Lax pairs}\nequation\label{laxsec}
In view of (\ref{qpositive}), we may define $q(x,t)$ by
\begin{equation}\label{qdef}
  q(x,t) = \bigl[u(x,t) - u_{xx}(x,t) + \kappa\bigr]^{\frac{1}{3}}, \qquad 0 \leq x < \infty, \quad 0 \leq t< T.
\end{equation}
For simplicity, we henceforth assume that $\kappa = 1$. 
Equation (\ref{DP}) admits the Lax pair (see \cite{BS2011})
\begin{equation}\label{psilax}
\begin{cases}
  \psi_x(x,t,k) = L(x,t,k) \psi(x,t,k), \\
  \psi_t(x,t,k) = Z(x,t,k) \psi(x,t,k),
\end{cases}
\end{equation}
where $k \in \hat{\C} = \C \cup \{\infty\}$ is a spectral parameter, $\psi(x,t, k)$ is a $3\times 3$-matrix valued eigenfunction, the $3\times 3$-matrix valued functions $L$ and $Z$ are defined by
$$L(x,t, k) = \begin{pmatrix} 0 & 1 & 0 \\ 0 & 0 & 1 \\ \lambda q^3 & 1 & 0 \end{pmatrix},
\qquad
Z(x,t, k) = \begin{pmatrix} u_x - \frac{2}{3\lambda} & - u & \frac{1}{\lambda} \\ u + 1 & \frac{1}{3\lambda} & -u \\
u_x - \lambda u q^3 & 1 & - u_x + \frac{1}{3\lambda} \end{pmatrix},$$
and, following \cite{CIL2010, BS2011}, we define $\lambda = \lambda(k)$Ê in terms of $k$ by
$$\lambda = \frac{1}{3\sqrt{3}}\biggl(k^3 + \frac{1}{k^3}\biggr).$$

The main difficulties of the spectral analysis are related to the singularities of $L$ and $Z$. These occur at the points where $\lambda = 0$ (i.e. at $k = e^{\frac{mi\pi}{3} - \frac{i\pi}{6}}$, $m = 1, \dots, 6$), where $Z$ is singular, and at the points where $\lambda = \infty$  (i.e. at $k = \infty$ and $k = 0$), where $L$ and $Z$ are singular. 
In order to formulate a Riemann-Hilbert problem, we will define transformed eigenfunctions $\Phi(x,t,k)$ and $\tilde{\Phi}(x,t,k)$ which are well-behaved near the points where $\lambda = 0$ and $\lambda = \infty$ respectively.

\subsection{Lax pair suitable near $\lambda = 0$}
In order to define eigenfunctions which are well-behaved near the points where $\lambda = 0$, i.e. near the points $\{K_j\}_1^6$ defined by
$$K_j = e^{\frac{\pi i j}{3} - \frac{i\pi}{6}}, \qquad j = 1, \dots, 6,$$ 
we transform the Lax pair (\ref{psilax}) as follows. We define $\{l_j\}$ and $\{z_j\}$ for $j = 1,2,3$ by
\begin{align}\label{lmexpressions}
&l_j(k) = \frac{1}{\sqrt{3}}\left(\omega^j k + \frac{1}{\omega^j k}\right), \quad
z_j(k) = \sqrt{3}\left(\frac{(\omega^j k)^2 + (\omega^j k)^{-2}}{k^3 + k^{-3}}\right), \quad \, k \in \C,
\end{align}
and let $\mathcal{L}$ and $\mathcal{Z}$Ê denote the corresponding diagonal matrices:
$$\mathcal{L} = \text{diag}(l_1 , l_2 , l_3 ), \qquad \mathcal{Z}  = \text{diag}(z_1 , z_2 , z_3 ).$$
The matrix-valued function Ê$P(k)$ defined by
\begin{align}\label{Pdef}
P(k) = \begin{pmatrix} 1 & 1 & 1 \\ l_1(k) & l_2(k) & l_3(k) \\ l_1^2(k) & l_2^2(k) & l_3^2(k) \end{pmatrix}, \qquad k \in \C,
\end{align}
diagonalizes the matrices
\begin{align}
L_\infty := \lim_{x \to \infty} L = \begin{pmatrix} 0 & 1 & 0 \\ 0 & 0 & 1 \\ \lambda & 1 & 0 \end{pmatrix}, \qquad
Z_\infty := 	\lim_{x \to \infty} Z = \begin{pmatrix} -\frac{2}{3\lambda} & 0 & \frac{1}{\lambda} \\ 1 & \frac{1}{3\lambda} & 0 \\ 0 & 1 & \frac{1}{3\lambda} \end{pmatrix},
\end{align}
as follows:
\begin{align}
L_\infty = P \mathcal{L} P^{-1},
\qquad
Z_\infty = 	P\mathcal{Z} P^{-1}.
\end{align}
Thus, the eigenfunction $\Phi$ introduced by
\begin{align}\label{Phidef}
  \psi(x,t,k) = P(k)\Phi(x,t,k) e^{\mathcal{L}(k) x + \mathcal{Z}(k) t},
\end{align}  
satisfies the Lax pair
\begin{equation}\label{Philax}
\begin{cases}
  \Phi_x - [\mathcal{L}, \Phi] = V_1 \Phi, \\
  \Phi_t - [\mathcal{Z}, \Phi] = V_2 \Phi,
\end{cases}
\end{equation}
where $V_1(x,t,k)$ and $V_2(x,t,k)$ are defined by
\begin{subequations}\label{V1V2def}
\begin{align}
& V_1 = P^{-1}\begin{pmatrix} 0 & 0 & 0 \\ 0 & 0 & 0 \\ \lambda (q^3 - 1) & 0 & 0 \end{pmatrix}P,
	\\
& V_2 = P^{-1}\begin{pmatrix} u_x & - u & 0 \\ u & 0 & -u \\ u_x - \lambda u q^3 & 0 & -u_x \end{pmatrix}P.
\end{align}
\end{subequations}
Indeed, (\ref{Philax}) follows from (\ref{psilax}), the definition (\ref{Phidef}) of $\Phi$, and the identities
$$L = P(\mathcal{L} + V_1)P^{-1}, \qquad Z = P(\mathcal{Z} + V_2)P^{-1}.$$

Let $K_j = e^{\frac{\pi i j}{3} - \frac{\pi i}{6}}$, $j = 1, \dots, 6$, denote the points where $\lambda = 0$ and let $\varkappa_j = e^{\frac{\pi i(j-1)}{3}}$, $j = 1, \dots, 6$, denote the sixth roots of unity, see Figure \ref{Kjs.pdf} below.
The definition (\ref{Phidef}) of $\Phi$ is chosen so that $V_1$ and $V_2$ have the properties stated in the following lemma.

\begin{lemma}\label{V1V2lemma}
  The functions $V_1$ and $V_2$ have the following properties:
  \begin{itemize} 
     \item As $k \to K_j$, 
 \begin{align*}
 &   V_1(x,t,k) = O(k - K_j), \qquad k \to K_j, \quad j = 1, \dots, 6.
\end{align*}

   \item The leading order term of $V_2(x,t,k)$ as $k \to K_j$ is off-diagonal:
\begin{align}\nonumber
&    V_2(x,t,k) = \mathcal{V}(x,t) + O(k - K_1), && k \to K_1,
	\\\nonumber
  & V_2(x,t,k) = \mathcal{A}^{-1} \mathcal{B}  \mathcal{V}(x,t) \mathcal{B}\mathcal{A} + O(k - K_2), && k \to K_2,
	\\ \nonumber
  & V_2(x,t,k) = \mathcal{A}^{-1}  \mathcal{V}(x,t) \mathcal{A} + O(k - K_3), && k \to K_3,
	\\\nonumber
  & V_2(x,t,k) = \mathcal{A} \mathcal{B}  \mathcal{V}(x,t)\mathcal{B}\mathcal{A}^{-1} + O(k - K_4), && k \to K_4,
	\\\nonumber
  & V_2(x,t,k) = \mathcal{A}  \mathcal{V}(x,t) \mathcal{A}^{-1} + O(k - K_5), && k \to K_5,
  	\\\label{V2nearkj}
  & V_2(x,t,k) = \mathcal{B}  \mathcal{V}(x,t) \mathcal{B} + O(k - K_6), && k \to K_6,
\end{align}
where the off-diagonal matrix $ \mathcal{V}(x,t)$ is defined by
\begin{align}\label{mathcalVdef}
&  \mathcal{V} = \begin{pmatrix} 0 & \frac{-u + u_x}{2} & 0 \\
u + u_x & 0 & -u + u_x \\
0 & \frac{u + u_x}{2} & 0 \end{pmatrix},
\end{align}
and
\begin{align}\label{ABdef}
\mathcal{A} =  \begin{pmatrix}
0 & 0 & 1 \\
 1 & 0 & 0 \\
 0 & 1 & 0\end{pmatrix}, \qquad
  \mathcal{B} =  \begin{pmatrix}
0 & 1 & 0 \\
1 & 0 & 0 \\
0 & 0 & 1 \end{pmatrix}.
\end{align}

 \item $V_1$ and $V_2$ are analytic for $k \in \hat{\C} \setminus \bigl(\{0, \infty\} \cup \{\varkappa_j\}_{j=1}^6\bigr)$.

  \item $V_1(x,t,k)$ and $V_2(x,t,k)$ decay to zero as $x \to \infty.$
  
\end{itemize}
\end{lemma}
\proofbegin
The behavior of the functions $V_1$ and $V_2$ near $\{K_j\}_1^6$ follows by Taylor series expansion. The stated analyticity properties follow from definition (\ref{V1V2def}) (the points $\{\varkappa_j\}_{1}^6$ must be excluded from the analyticity domain because $P^{-1}$ has poles at these points).
The behavior as $x \to \infty$ is immediate from (\ref{V1V2def}). 
\proofend

\subsection{Lax pair suitable near $\lambda = \infty$}
In order to define eigenfunctions which are well-behaved near the points where $\lambda = \infty$, i.e. near $k = \infty$ and $k = 0$, we transform the Lax pair (\ref{psilax}) as follows cf. \cite{BS2011}. 
Define $y(x,t)$ by
\begin{equation}\label{ydef}
  y(x,t) = \int_{(0,0)}^{(x,t)} q(x',t')\left(dx' - u(x', t')dt'\right),
\end{equation}  
and introduce the eigenfunction $\tilde{\Phi}$ by
\begin{align}\label{tildePhidef}
  \psi(x,t,k) = D(x,t) P(k) \tilde{\Phi}(x,t,k) e^{\mathcal{L}(k)y(x,t) + \mathcal{Z}(k) t},
\end{align}
where
\begin{align}\label{Ddef}
  D(x,t) = \begin{pmatrix} \frac{1}{q(x,t)} & 0 & 0 \\ 0 & 1 & 0 \\ 0 & 0 & q(x,t) \end{pmatrix}.
\end{align}  
The function $y(x,t)$ is well-defined by (\ref{ydef}), because the conservation law
\begin{align}\label{conslaw}
  q_t + \left(q u\right)_x = 0,
\end{align}  
implies that the integral in (\ref{ydef}) is independent of the path of integration.

The eigenfunction $\tilde{\Phi}$ satisfies the Lax pair
\begin{equation}\label{tildePhilax}
\begin{cases}
  \tilde{\Phi}_x - [q\mathcal{L}, \tilde{\Phi}] = \tilde{V}_1 \tilde{\Phi}, \\
  \tilde{\Phi}_t - [\mathcal{Z} - uq\mathcal{L}, \tilde{\Phi}] = \tilde{V}_2 \tilde{\Phi},
\end{cases}
\end{equation}
where $\tilde{V}_1(x,t,k)$ and $\tilde{V}_2(x,t,k)$ are defined by
\begin{subequations}\label{tildeV1V2def}
\begin{align}
& \tilde{V}_1 = P^{-1}\begin{pmatrix} \frac{q_x}{q} & 0 & 0 \\ 0 & 0 & 0 \\ 0 & \frac{1}{q} - q & - \frac{q_x}{q} \end{pmatrix}P,
	\\
& \tilde{V}_2 = P^{-1} \left[\begin{pmatrix} -\frac{uq_x}{q} & 0 & 0 \\ \frac{u + 1}{q} - 1 & 0 & 0 \\ \frac{u_x}{q^2} & \frac{1}{q} - 1 + uq & \frac{uq_x}{q} \end{pmatrix} + \frac{q^2 - 1}{\lambda}\begin{pmatrix} 0 & 0 & 1 \\ 0 & 0 & 0 \\ 0 & 0 & 0 \end{pmatrix}\right] P.
\end{align}
\end{subequations}
Indeed, (\ref{tildePhilax}) follows from (\ref{psilax}), the definition (\ref{tildePhidef}) of $\tilde{\Phi}$, and the identities
\begin{align*}
& -D^{-1}D_x + D^{-1}LD = P(q\mathcal{L} + \tilde{V}_1)P^{-1}, 
	\\
& -D^{-1}D_t + D^{-1}ZD = P(\mathcal{Z} - uq \mathcal{L} + \tilde{V}_2)P^{-1}.
\end{align*}

\begin{remark}\upshape
We can use (\ref{conslaw}) to express $q_x$ on the right-hand side of (\ref{tildeV1V2def}) in terms of $q_t, u, u_x$, and $u_{xx}$. In particular, the function $\tilde{V}_2$ can be defined on the boundary $\{x = 0, 0 \leq t < T\}$ in terms of $g_0(t)$, $g_1(t)$, and $g_2(t)$ alone.
\end{remark}

The definition (\ref{tildePhidef}) of $\tilde{\Phi}$ is chosen so that $\tilde{V}_1$ and $\tilde{V}_2$ have the properties stated in the following lemma.

\begin{lemma}\label{tildeV1V2lemma}
  The functions $\tilde{V}_1(x,t,k)$ and $\tilde{V}_2(x,t,k)$ defined in (\ref{tildeV1V2def}) have the following properties:
  \begin{itemize}
   \item The leading order terms of $\tilde{V}_1$ and $\tilde{V}_2$ as $k \to \infty$ are off-diagonal:
    \begin{align} 
 \tilde{V}_1 = \tilde{\mathcal{V}} + O(1/k), \qquad  \tilde{V}_2 = -u \tilde{\mathcal{V}} + O(1/k), \qquad k \to \infty,
\end{align}
where the off-diagonal matrix $\tilde{\mathcal{V}}$ is given by
\begin{align*}
& \tilde{\mathcal{V}} = \frac{q_x}{\omega (1-\omega) q} \begin{pmatrix} 0 & \omega & 1+ \omega \\
1 + \omega & 0 & \omega \\
\omega & 1 +\omega & 0 \end{pmatrix}.
\end{align*}

   \item The leading order terms of $\tilde{V}_1$ and $\tilde{V}_2$ as $k \to 0$ are off-diagonal:
    \begin{align*}
 \tilde{V}_1 = \mathcal{B}\tilde{\mathcal{V}}\mathcal{B} + O(k), \qquad  \tilde{V}_2 = -u \mathcal{B}\tilde{\mathcal{V}}\mathcal{B} + O(k), \qquad k \to 0,
\end{align*}
where $\mathcal{B}$ is defined in (\ref{ABdef}).

 \item $\tilde{V}_1$ and $\tilde{V}_2$ are analytic for $k \in \hat{\C} \setminus \{K_j, \varkappa_j\}_{j = 1}^{6}$.
  
  \item $\tilde{V}_1(x,t,k)$ and $\tilde{V}_2(x,t,k)$ decay to zero as $x \to \infty.$
\end{itemize}
\end{lemma}
\proofbegin
The behavior of the functions $\tilde{V}_1$ and $\tilde{V}_2$ near $k = \infty$ and $k = 0$ follows by Taylor series expansion. The stated analyticity properties then follow from definition (\ref{tildeV1V2def}) (the points $\{\varkappa_j\}_{1}^6$ must be excluded from the analyticity domain because $P^{-1}$ has poles at these points). The behavior as $x \to \infty$ is immediate from (\ref{tildeV1V2def}). 
\proofend

\subsection{Symmetries}
The expansions in Lemmas \ref{V1V2lemma} and \ref{tildeV1V2lemma} are consistent with the following symmetry properties satisfied by the Lax pairs (\ref{Philax}) and (\ref{tildePhilax}).

\begin{lemma}\label{symmlemma} 
Let $F$ denote one of the $3 \times 3$-matrix valued functions $\mathcal{L}$, $\mathcal{Z}$, $V_1$, $V_2$, $\tilde{V}_1$, or $\tilde{V}_2$. Then $F$ obeys the $\Z_3$ symmetry
\begin{align}
&  F(k) = \mathcal{A} F(\omega k)\mathcal{A}^{-1}, \qquad k \in \C, 
\end{align}
as well as the $\Z_2$ symmetries
\begin{align}
& F(k) = \mathcal{B} \overline{F(\overline{k})}\mathcal{B},	\qquad
F(k) = \mathcal{B} F(1/k)\mathcal{B}, \qquad k \in \C,
\end{align}
where $\mathcal{A}$, $\mathcal{B}$ are defined in (\ref{ABdef}) and we have suppressed the $(x,t)$-dependence.
\end{lemma}
\proofbegin
This is a consequence of the definitions and the following symmetry properties of $P(k)$:
\begin{align}\label{Psymmetries}
& P(k) = P(\omega k)\mathcal{A}^{-1}, \quad
P(k) =  \overline{P(\overline{k})}\mathcal{B},\quad
P(k) = P(1/k)\mathcal{B}, \qquad k \in \C.
\end{align}
\proofend

\section{Analytic eigenfunctions}\nequation\label{eigensec}
In this section we define eigenfunctions of the Lax pairs (\ref{Philax}) and (\ref{tildePhilax}) in terms of linear integral equations. These eigenfunctions are the basic building blocks needed for the formulation of a Riemann-Hilbert problem. We first use the Lax pair (\ref{Philax}) to define eigenfunctions which are well-behaved near the points $K_j = e^{\frac{\pi i j}{3} - \frac{\pi i}{6}}$, $j = 1, \dots, 6$. 
We then use the Lax pair (\ref{tildePhilax}) to define eigenfunctions which are well-behaved near the points $k = \infty$ and $k = 0$. 
Together these two sets of eigenfunctions will be used to define a sectionally meromorphic function suitable for the formulation of a RH problem.

Let equation (\ref{DP}) be valid in the half-line domain (\ref{halflinedomain}). For a diagonal matrix $D$, we introduce the notation $\hat{D}$ for the operator which acts on a matrix $A$ by $\hat{D}A = [D, A]$; in particular $e^{\hat{D}}A = e^D A e^{-D}$. 

\subsection{The first set of eigenfunctions}
\begin{figure}
\begin{center}
    \includegraphics[width=.3\textwidth]{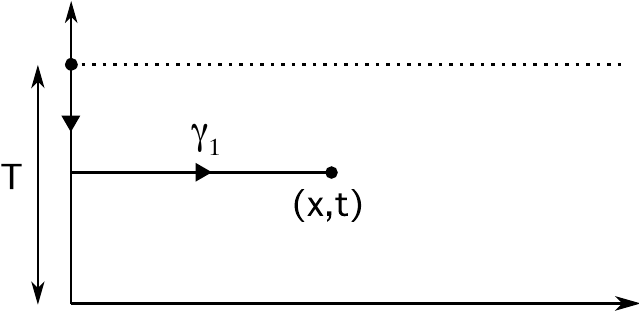} \quad
    \includegraphics[width=.3\textwidth]{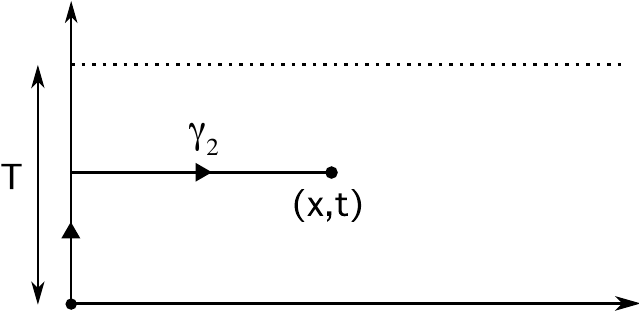} \quad
    \includegraphics[width=.3\textwidth]{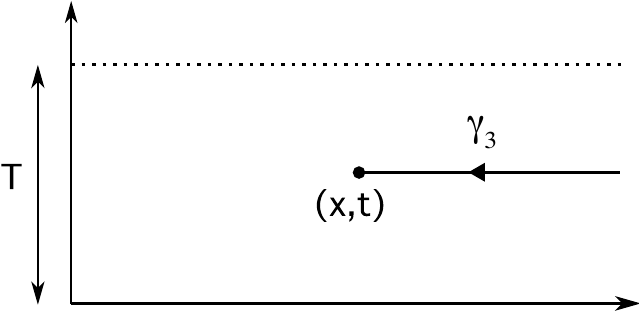} \\
     \begin{figuretext}\label{mucontours.pdf}
       The contours $\gamma_1$, $\gamma_2$, and $\gamma_3$ in the $(x, t)$-plane.
     \end{figuretext}
     \end{center}
\end{figure}

\begin{figure}
\begin{center}
 \begin{overpic}[width=.8\textwidth]{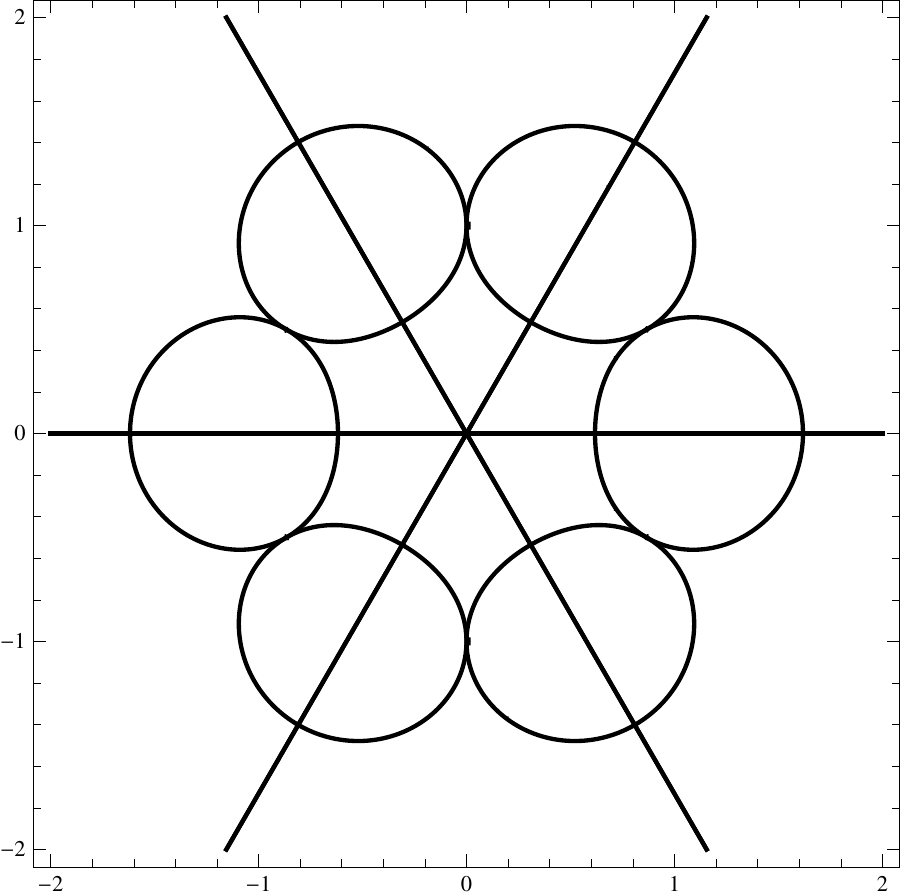}
      \put(86,72){$D_1$}
      \put(59,55){$D_1$}
      \put(50,90){$D_2$}
      \put(50,61){$D_2$}
     \put(14,72){$D_3$}
      \put(40,55){$D_3$}
     \put(14,28){$D_4$}
      \put(40,45){$D_4$}
      \put(50,10){$D_5$}
      \put(50,39){$D_5$}
      \put(86,28){$D_6$}
      \put(59,45){$D_6$}
      \put(74,56){$D_7$}
      \put(67,68){$D_8$}
      \put(56,74){$D_9$}
      \put(42,74){$D_{10}$}
      \put(32,68){$D_{11}$}
      \put(25,56){$D_{12}$}
      \put(25,44){$D_{13}$}
      \put(32,32){$D_{14}$}
      \put(42,26){$D_{15}$}
      \put(56,26){$D_{16}$}
      \put(67,32){$D_{17}$}
      \put(74,44){$D_{18}$}
 \end{overpic}
    \qquad \qquad
     \begin{figuretext}\label{Dns.pdf}
       The sets $D_n$, $n = 1, \dots, 18$, which decompose the complex $k$-plane. 
         \end{figuretext}
     \end{center}
\end{figure}

The first set of eigenfunctions is defined using the Lax pair (\ref{Philax}). We write (\ref{Philax}) in differential form as
\begin{equation}\label{Psilaxdiffform}
d\left(e^{-\hat{\mathcal{L}} x - \hat{\mathcal{Z}} t} \Phi \right) = W,
\end{equation}
where the closed one-form $W(x,t,k)$ is defined by
\begin{equation}\label{Wdef}  
  W = e^{-\hat{\mathcal{L}} x - \hat{\mathcal{Z}} t}(V_1 dx + V_2 dt) \Phi.
\end{equation}  
We define three contours $\{\gamma_j\}_1^3$ in the $(x, t)$-plane going from $(x_j, t_j)$ to $(x,t)$, where $(x_1, t_1) = (0, T)$, $(x_2, t_2) = (0, 0)$, and $(x_3, t_3) = (\infty, t)$; we choose the particular contours shown in Figure \ref{mucontours.pdf}.
We define eighteen open, pairwisely disjoint subsets $\{D_n\}_1^{18}$ of the Riemann $k$-sphere by (see Figure \ref{Dns.pdf})
{\allowdisplaybreaks
\begin{align*}
&D_1 = \{k \in \hat{\C}\,|\, \text{Re}\,l_1 < \text{Re}\,l_2 < \text{Re}\,l_3 \text{  and  } 
\text{Re}\,z_1 < \text{Re}\,z_2 < \text{Re}\,z_3 \},
	\\
&D_2 = \{k \in \hat{\C}\,|\, \text{Re}\,l_1 < \text{Re}\,l_3 < \text{Re}\,l_2 \text{  and  } 
\text{Re}\,z_1 < \text{Re}\,z_3 < \text{Re}\,z_2 \},
	\\
&D_3 = \{k \in \hat{\C}\,|\, \text{Re}\,l_3 < \text{Re}\,l_1 < \text{Re}\,l_2 \text{  and  } 
\text{Re}\,z_3 < \text{Re}\,z_1 < \text{Re}\,z_2 \},
	\\
&D_4 = \{k \in \hat{\C}\,|\, \text{Re}\,l_3 < \text{Re}\,l_2 < \text{Re}\,l_1 \text{  and  } 
\text{Re}\,z_3 < \text{Re}\,z_2 < \text{Re}\,z_1 \},
	\\
&D_5 = \{k \in \hat{\C}\,|\, \text{Re}\,l_2 < \text{Re}\,l_3 < \text{Re}\,l_1 \text{  and  } 
\text{Re}\,z_2 < \text{Re}\,z_3 < \text{Re}\,z_1 \},
	\\
&D_6 = \{k \in \hat{\C}\,|\, \text{Re}\,l_2 < \text{Re}\,l_1 < \text{Re}\,l_3 \text{  and  } 
\text{Re}\,z_2 < \text{Re}\,z_1 < \text{Re}\,z_3 \},
	\\
&D_7 = \{k \in \hat{\C}\,|\, \text{Re}\,l_1 < \text{Re}\,l_2 < \text{Re}\,l_3 \text{  and  } 
\text{Re}\,z_2 < \text{Re}\,z_1 < \text{Re}\,z_3 \},
	\\
&D_8 = \{k \in \hat{\C}\,|\, \text{Re}\,l_1 < \text{Re}\,l_2 < \text{Re}\,l_3 \text{  and  } 
\text{Re}\,z_1 < \text{Re}\,z_3 < \text{Re}\,z_2 \},
	\\
&D_9 = \{k \in \hat{\C}\,|\, \text{Re}\,l_1 < \text{Re}\,l_3 < \text{Re}\,l_2 \text{  and  } 
\text{Re}\,z_1 < \text{Re}\,z_2 < \text{Re}\,z_3 \},
	\\
&D_{10} = \{k \in \hat{\C}\,|\, \text{Re}\,l_1 < \text{Re}\,l_3 < \text{Re}\,l_2 \text{  and  } 
\text{Re}\,z_3 < \text{Re}\,z_1 < \text{Re}\,z_2 \},
	\\
&D_{11} = \{k \in \hat{\C}\,|\, \text{Re}\,l_3 < \text{Re}\,l_1 < \text{Re}\,l_2 \text{  and  } 
\text{Re}\,z_1 < \text{Re}\,z_3 < \text{Re}\,z_2 \},
	\\
&D_{12} = \{k \in \hat{\C}\,|\, \text{Re}\,l_3 < \text{Re}\,l_1 < \text{Re}\,l_2 \text{  and  } 
\text{Re}\,z_3 < \text{Re}\,z_2 < \text{Re}\,z_1 \},
	\\
&D_{13} = \{k \in \hat{\C}\,|\, \text{Re}\,l_3 < \text{Re}\,l_2 < \text{Re}\,l_1 \text{  and  } 
\text{Re}\,z_3 < \text{Re}\,z_1 < \text{Re}\,z_2 \},
	\\
&D_{14} = \{k \in \hat{\C}\,|\, \text{Re}\,l_3 < \text{Re}\,l_2 < \text{Re}\,l_1 \text{  and  } 
\text{Re}\,z_2 < \text{Re}\,z_3 < \text{Re}\,z_1 \},
	\\
&D_{15} = \{k \in \hat{\C}\,|\, \text{Re}\,l_2 < \text{Re}\,l_3 < \text{Re}\,l_1 \text{  and  } 
\text{Re}\,z_3 < \text{Re}\,z_2 < \text{Re}\,z_1 \},
	\\
&D_{16} = \{k \in \hat{\C}\,|\, \text{Re}\,l_2 < \text{Re}\,l_3 < \text{Re}\,l_1 \text{  and  } 
\text{Re}\,z_2 < \text{Re}\,z_1 < \text{Re}\,z_3 \},
	\\
&D_{17} = \{k \in \hat{\C}\,|\, \text{Re}\,l_2 < \text{Re}\,l_1 < \text{Re}\,l_3 \text{  and  } 
\text{Re}\,z_2 < \text{Re}\,z_3 < \text{Re}\,z_1 \},
	\\
&D_{18} = \{k \in \hat{\C}\,|\, \text{Re}\,l_2 < \text{Re}\,l_1 < \text{Re}\,l_3 \text{  and  } 
\text{Re}\,z_1 < \text{Re}\,z_2 < \text{Re}\,z_3 \}.
	\\
\end{align*}
}
For each $n = 1, \dots, 18$, we define a solution $\Phi_n(x,t,k)$ of (\ref{Philax}) by the following system of integral equations: 
\begin{equation}\label{Phinintegraleq}
(\Phi_n)_{ij}(x,t,k) = \delta_{ij} + \int_{\gamma_{ij}^n} \left(e^{\hat{\mathcal{L}}(k) x + \hat{\mathcal{Z}}(k) t} W_n(x',t',k)\right)_{ij}, \qquad k \in D_n, \quad i,j = 1, 2,3,
\end{equation}
where the contours $\gamma^n_{ij}$, $n = 1, \dots, 18$, $i,j = 1, 2,3$, are defined by
 \begin{align} \label{gammaijnudef}
 \gamma_{ij}^n =  \begin{cases}
 \gamma_1,  \qquad \text{Re}\, l_i(k) < \text{Re}\, l_j(k), \quad \text{Re}\, z_i(k) \geq \text{Re}\, z_j(k),
	\\
\gamma_2,  \qquad \text{Re}\, l_i(k) < \text{Re}\, l_j(k),\quad \text{Re}\, z_i(k) < \text{Re}\, z_j(k),
	\\
\gamma_3,  \qquad \text{Re}\, l_i(k) \geq \text{Re}\, l_j(k),
	\\
\end{cases} \quad \text{for} \quad k \in D_n,
\end{align}
and $W_n$ is given by (\ref{Wdef}) with $\Phi$ replaced by $\Phi_n$.

\begin{proposition}\label{Phiprop}
For each $n = 1, \dots, 18$, the function $\Phi_n(x,t,k)$ is well-defined by equation (\ref{Phinintegraleq}) for $k \in \bar{D}_n$ and $(x,t)$ in the domain (\ref{halflinedomain}). For any fixed point $(x,t)$, $\Phi_n$ is bounded and analytic as a function of $k \in D_n$ away from the points $\{\infty, 0\} \cup \{\varkappa_j\}_1^6 \cup \{k_j\}$, where $\{k_j\}$ denotes a possibly empty discrete set of singularities at which the Fredholm determinant vanishes. Moreover, $\Phi_n$ admits a bounded and continuous extension to $\bar{D}_n$ away from these points.
\end{proposition}
\proofbegin
The definition of $\{\gamma_j\}_1^3$ implies the following relations on the contours:
\begin{align}\nonumber
& \gamma_1: x - x' \geq 0, \qquad t - t' \leq 0,
	\\ \label{contourrelations}
& \gamma_2: x - x' \geq 0, \qquad t - t' \geq 0,
	\\ \nonumber
& \gamma_3: x - x' \leq 0, \qquad t - t' = 0.
\end{align}
The $(ij)$th entry of the integral equation (\ref{Phinintegraleq}) involves the exponential factor 
\begin{equation}\nonumber
e^{(l_i - l_j)(x - x') + (z_i - z_j)(t - t')}.
\end{equation}
In view of (\ref{gammaijnudef}) and (\ref{contourrelations}) this factor remains bounded for $k \in D_n$ when integrated along the contour $\gamma_{ij}^n$. The decay of $V_1$ and $V_2$ as $x \to \infty$ together with the analyticity properties of these functions now implies that the solution $\Phi_n$ exists and that it has the stated propertes---a proof using an extension of the standard Fredholm theory is given in Appendix B of \cite{L2012}.
\proofend

The symmetries of Lemma \ref{symmlemma} imply corresponding symmetry properties for the eigenfunctions $\{\Phi_n\}_1^{18}$.

\begin{figure}
\begin{center}
 \begin{overpic}[width=.65\textwidth]{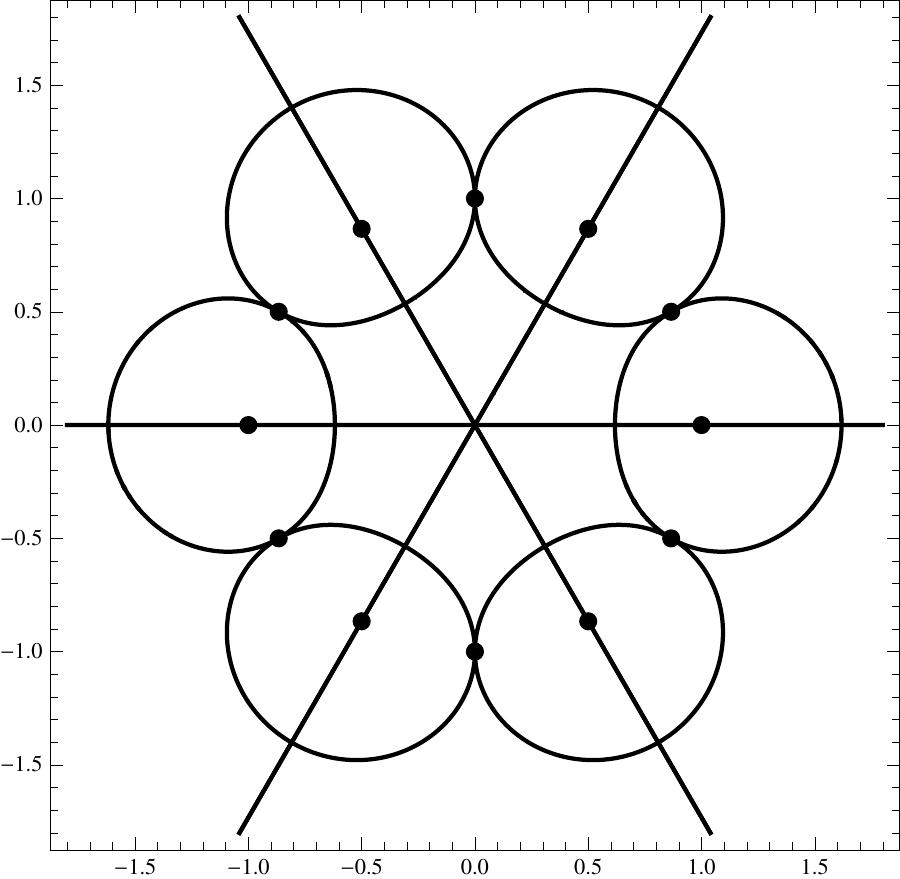}
      \put(75,59){\small $K_1$}
      \put(54.5,74.5){\small $K_2$}
      \put(31,64.5){\small $K_3$}
      \put(26,40){\small $K_4$}
      \put(46,24){\small $K_5$}
      \put(70,34){\small $K_6$}
            \put(75,46.5){ $\varkappa_1$}
      \put(65,70){ $\varkappa_2$}
      \put(40,73){ $\varkappa_3$}
      \put(24,53){ $\varkappa_4$}
      \put(33,29){ $\varkappa_5$}
      \put(59,26){ $\varkappa_6$}
 \end{overpic}
    \qquad \qquad
     \begin{figuretext}\label{Kjs.pdf}
       The six points $K_j = e^{\frac{\pi i j}{3} - \frac{\pi i}{6}}$, $j = 1, \dots, 6$, where $\lambda = 0$, and the six points $\varkappa_j = e^{\frac{\pi i(j-1)}{3}}$, $j = 1, \dots, 6$, where $P^{-1}(k)$ has poles.
            \end{figuretext}
     \end{center}
\end{figure}
\begin{lemma}\label{Phinsymmlemma} 
Define the sectionally meromorphic function $\Phi_*(x,t,k)$ by
$$\Phi_*(x,t,k) = \Phi_n(x,t,k), \qquad k \in D_n.$$
Then
\begin{align}\label{Phisymmetries}
&  \Phi_*(k) = \mathcal{A} \Phi_*(\omega k)\mathcal{A}^{-1}, \quad  \Phi_*(k) = \mathcal{B} \overline{\Phi_*(\overline{k})}\mathcal{B},	\quad
\Phi_*(k) = \mathcal{B} \Phi_*(1/k)\mathcal{B}, \quad k \in \C,
\end{align}
where $\mathcal{A}$, $\mathcal{B}$ are defined in (\ref{ABdef}) and we have suppressed the $(x,t)$-dependence.
\end{lemma}
\proofbegin
We will prove the first symmetry property in the case when $k \in D_1$, that is, we will prove that
\begin{align}\label{Phi1Phi3}
  \Phi_1(x,t,k) = \mathcal{A} \Phi_3(x,t,\omega k)\mathcal{A}^{-1}, \qquad k \in D_1.
\end{align}  
Define $\phi_1(x,t,k)$ and $\phi_3(x,t,k)$ by
$$\phi_1 = \Phi_1 e^{\mathcal{L}x + \mathcal{Z} t}, \qquad \phi_3 = \Phi_3 e^{\mathcal{L}x + \mathcal{Z} t}.$$
The symmetries of Lemma \ref{symmlemma} imply that the equations
$$\begin{cases}
\phi_x(x,t,k) - \mathcal{L}(k) \phi(x,t,k) = V_1(x,t,k) \phi(x,t,k), \\
\phi_t(x,t,k) - \mathcal{L}(k) \phi(x,t,k) = V_2(x,t,k) \phi(x,t,k),
\end{cases}$$
hold both for $\phi(x,t,k) = \phi_1(x,t,k)$ and for $\phi(x,t,k) = \mathcal{A} \phi_3(x,t,\omega k) \mathcal{A}^{-1}$.
Therefore, the $3Ê\times 3$-matrix valued function $J(k)$ defined by
%$$\Phi_1(x,t,k)e^{\mathcal{L}x + \mathcal{Z} t} = \mathcal{A} \Phi_3(x,t,\omega k)e^{\mathcal{L}x + \mathcal{Z} t} \mathcal{A}^{-1} J(k), \qquad k \in D_1.$$
\begin{align}\label{Jdef}
  J(k) = \mathcal{A} e^{-\mathcal{L}(\omega k) x - \mathcal{Z}(\omega k) t}\Phi_3^{-1}(x,t,\omega k) \mathcal{A}^{-1} \Phi_1(x,t,k)e^{\mathcal{L}(k)x + \mathcal{Z}(k) t}, \qquad k \in D_1,
\end{align}
is independent of $(x,t)$. Since $\mathcal{A} e^{-\mathcal{L}(\omega k) x - \mathcal{Z}(\omega k) t} = e^{-\mathcal{L}( k) x - \mathcal{Z}( k) t} \mathcal{A}$, it only remains to prove that $J(k) = I$.
By (\ref{gammaijnudef}), the matrices $\gamma^1$ and $\gamma^3$ defined by $(\gamma^1)_{ij} := \gamma_{ij}^{1}$ and $(\gamma^3)_{ij} := \gamma_{ij}^{3}$ are given by
\begin{align}\label{gamma1gamma3}
\gamma^1 = \begin{pmatrix} 
\gamma_3 & \gamma_2 & \gamma_2 \\
\gamma_3 & \gamma_3 & \gamma_2 \\
\gamma_3 & \gamma_3 & \gamma_3 \end{pmatrix}, \qquad
\gamma^3 = \begin{pmatrix} 
\gamma_3 & \gamma_2 & \gamma_3 \\
\gamma_3 & \gamma_3 & \gamma_3 \\
\gamma_2 & \gamma_2 & \gamma_3 \end{pmatrix}.
\end{align}
Consequently, 
\begin{align}\label{Phi1Phi3at00}
\Phi_1(0,0,k)
= 
\begin{pmatrix} * & 0 & 0 \\
* & * & 0 \\
* & * & * 
\end{pmatrix}, \qquad 
\Phi_3(0,0,\omega k)
= 
\begin{pmatrix} * & 0 & * \\
* & * & * \\
0 & 0 & * 
\end{pmatrix}, \qquad k \in D_1,
\end{align}
where $*$ denotes an unspecified entry. Evaluating (\ref{Jdef}) at $(x,t) = (0,0)$ and using (\ref{Phi1Phi3at00}) as well as the determinant condition $\det \Phi_3 = 1$, we find that $J(k)$ has the form
\begin{align}\label{Jstar1}
J(k) =  \begin{pmatrix} * & 0 & 0 \\
* & * & 0 \\
* & * & * 
\end{pmatrix}.
\end{align}
Similarly, evaluating (\ref{Phi1Phi3at00}) as $x \to\infty$ and using that
$$\lim_{x \to \infty} \Phi_1(x,0,k)
= 
\begin{pmatrix} 1 & * & * \\
0 & 1 & * \\
0 & 0 & 1 
\end{pmatrix}, \qquad 
\lim_{x \to \infty} \Phi_3(x,0,\omega k)
= 
\begin{pmatrix} 1 & * & 0 \\
0 & 1 & 0 \\
* & * & 1 
\end{pmatrix}, \qquad k \in D_1,
$$
we find that $J(k)$ has the form
\begin{align}\label{Jstar2}
J(k) =  \begin{pmatrix} 1 & * & * \\
0 & 1 & * \\
0 & 0 & 1
\end{pmatrix}.
\end{align}
Equations (\ref{Jstar1}) and (\ref{Jstar2}) show that $J(k) = I$. This completes the proof of (\ref{Phi1Phi3}). The other symmetries can be proved in a similar way. 
\proofend

The following lemma shows that $\Phi_nÊ\to I$ as $k \to K_j = e^{\frac{\pi i j}{3} - \frac{\pi i}{6}}$, $j = 1, \dots, 6$.

\begin{lemma}\label{Phinasymptoticslemma}
The eigenfunctions $\{\Phi_n(x,t,k)\}_{1}^{18}$ defined by the integral equation (\ref{Phinintegraleq}) satisfy 
\begin{equation}\label{Phinatlambda0}
 \Phi_n(x,t,k) = I + O(k - K_j) \quad \text{as} \quad k \to K_j, \quad k \in D_n, \quad j = 1, \dots, 6.
\end{equation}
\end{lemma}
\proofbegin
We substitute the expansion
$$\Phi(x,t,k) = \Phi^{(0)}(x,t) + \Phi^{(1)}(x,t)(k - K_1) + \dots, \qquad k \to K_1,$$
where the $\Phi^{(j)}$'s are independent of $k$, into the $t$-part of the Lax pair (\ref{Philax}). 
Using (\ref{V2nearkj}) and the fact that
$$\mathcal{Z} = -\frac{\omega}{2\sqrt{3}}\begin{pmatrix} 1 & 0 & 0 \\ 0 & -2 & 0 \\ 0 & 0 & 1 \end{pmatrix}  \frac{1}{k - K_1} 
+ \begin{pmatrix} \frac{-13 - 2\omega}{12} & 0 & 0 \\ 0 & \frac{1 + 2\omega}{6} & 0 \\ 0 & 0 & \frac{11  - 2\omega}{12} \end{pmatrix} + O(k - K_1), \quad k \to K_1,$$
we find that the terms of $O(1/(k -K_1))$ yield
$$\frac{\omega}{2\sqrt{3}}\left[\begin{pmatrix} 1 & 0 & 0 \\ 0 & -2 & 0 \\ 0 & 0 & 1 \end{pmatrix} , \Phi^{(0)}\right] = 0 \qquad \text{i.e.} \qquad \Phi^{(0)} = \begin{pmatrix} \Phi^{(0)}_{11} & 0 & \Phi^{(0)}_{13} \\ 0 & \Phi^{(0)}_{22} & 0 \\ \Phi^{(0)}_{31} & 0 & \Phi^{(0)}_{33} \end{pmatrix},$$
while the terms of $O(1)$ yield
\begin{align}\label{order1}
\Phi_t^{(0)} + \frac{\omega}{2\sqrt{3}}\left[\begin{pmatrix} 1 & 0 & 0 \\ 0 & -2 & 0 \\ 0 & 0 & 1 \end{pmatrix} , \Phi^{(1)}\right]
- \left[ \begin{pmatrix} \frac{-13 - 2\omega}{12} & 0 & 0 \\ 0 & \frac{1 + 2\omega}{6} & 0 \\ 0 & 0 & \frac{11  - 2\omega}{12} \end{pmatrix}, \Phi^{(0)}\right] = \mathcal{V} \Phi^{(0)},
\end{align}
where the off-diagonal matrix $\mathcal{V}$ is defined in (\ref{mathcalVdef}).
The diagonal terms of (\ref{order1}) imply that
$$\Phi^{(0)}_{11t} = \Phi^{(0)}_{22t}=\Phi^{(0)}_{33t} = 0,$$
whereas the (13) and (31) entries of (\ref{order1}) imply that
$$\Phi^{(0)}_{13t} + 2 \Phi^{(0)}_{13} = 0, \qquad \Phi^{(0)}_{31t} - 2 \Phi^{(0)}_{31} = 0.$$

On the other hand, the terms of $O(1)$ of the $x$-part of (\ref{Philax}) show that
$$\Phi^{(0)}_{11x} = \Phi^{(0)}_{22x} = \Phi^{(0)}_{33x} = 0, \qquad \Phi^{(0)}_{13x} + 2\Phi^{(0)}_{13} = 0, \qquad \Phi^{(0)}_{31x} - 2\Phi^{(0)}_{31} = 0.$$
Thus,
$$\Phi^{(0)} = \begin{pmatrix} c_1 & 0 & c_4 e^{-2x - 2t} \\ 0 & c_2 & 0 \\ c_5 e^{2x+2t} & 0 & c_3 \end{pmatrix},$$
where $c_j$, $j = 1, \dots, 5$, are constants independent of $x$ and $t$. 
For the eigenfunctions $\{\Phi_n\}_{1}^{18}$ defined by (\ref{Phinintegraleq}), evaluation at the points $(x,t) = (x_j, t_j)$, $j = 1,2,3$, implies that $c_1 = c_2 = c_3 = 1$ and $c_4 = c_5 = 0$. This proves (\ref{Phinatlambda0}) for $j = 1$. 
\proofend

The next lemma establishes the singularity structure of $\Phi$ near the points 
$\varkappa_j = e^{\frac{\pi i (j-1)}{3}}$, $j = 1, \dots, 6$, where $P^{-1}$ has simple poles.

\begin{lemma}\label{polestructurelemma}
The functions $\Phi_7$ and $\Phi_{8}$ satisfy
\begin{align}\label{Phi7at1}
&\Phi_7(x,t,k) = \frac{F(x,t)}{k-1}+ O(1), \qquad k \to 1, \quad k \in D_7,
	\\ \label{Phi8at1}
&\Phi_{8}(x,t,k) = \frac{G(x,t)}{k-1}+ O(1), \qquad k \to e^{\frac{i\pi}{3}}, \quad k \in D_8.
\end{align}
where the $3 \times 3$-matrix valued functions $F$ and $G$ have the form
\begin{align*}
& F(x,t) = \begin{pmatrix} f_1(x,t) & f_2(x,t) & f_3(x,t) \\
-f_1(x,t) & -f_2(x,t) & -f_3(x,t) \\
0 & 0 & 0 \end{pmatrix},
	\\
& G(x,t) = \begin{pmatrix} 
0 & 0 & 0  \\
g_1(x,t) & g_2(x,t) & g_3(x,t) \\
-g_1(x,t) & -g_2(x,t) & -g_3(x,t) \end{pmatrix},
\end{align*}
and $\{f_j(x,t), g_j(x,t)\}_1^3$ are scalar-valued functions.
\end{lemma}
\proofbegin
The behavior of $P^{-1}(k)$ near $k = 1$ is given by
\begin{align}\label{invPat1}
 P^{-1}(k) = \frac{i}{6\sqrt{3}}\begin{pmatrix} -2 & - \sqrt{3} & 3 \\ 2 & \sqrt{3} & -3 \\Ê0 & 0 & 0 
 \end{pmatrix}\frac{1}{k-1} + O(1), \qquad k \to 1.
\end{align} 
In view of (\ref{Phidef}), the function $\psi_7$ defined by
\begin{align}\label{psi9def}
  \Phi_7(x,t,k) = P^{-1}(k) \psi_7(x,t,k) e^{-\mathcal{L}(k) x - \mathcal{Z}(k) t}, \qquad k \in D_7,
\end{align}  
satisfies the Lax pair equations (\ref{psilax}), which are nonsingular atÊ $k = 1$. Hence $\psi_7$ is analytic at $k = 1$. Since the exponential $e^{-\mathcal{L}(k) x - \mathcal{Z}(k) t}$ is also analytic at $k = 1$, the expansion (\ref{Phi7at1}) follows immediately from (\ref{invPat1}) and (\ref{psi9def}).
The proof of (\ref{Phi8at1}) is similar.
\proofend

The behavior of the $\Phi_n$'s as $k$ approaches any of the $\varkappa_j$'s, follows from Lemma \ref{polestructurelemma} together with the symmetries of Lemma \ref{Phinsymmlemma}. For example, taking the Schwartz conjugate of (\ref{Phi7at1}) and using (\ref{Phisymmetries}), we find
$$\Phi_{18}(x,t,k) = \frac{\mathcal{B}\bar{F}(x,t)\mathcal{B}}{k-1}+ O(1), \qquad k \to 1, \quad k \in D_{18}.$$

\subsection{The second set of eigenfunctions}
The second set of eigenfunctions is defined using the Lax pair (\ref{tildePhilax}). We write (\ref{tildePhilax}) in differential form as
\begin{equation}\label{tildePhilaxdiffform}
d\left(e^{-\hat{\mathcal{L}} y - \hat{\mathcal{Z}} t} \tilde{\Phi} \right) = \tilde{W},
\end{equation}
where the closed one-form $\tilde{W}(x,t,k)$ is defined by
\begin{equation}\label{tildeWdef}  
  \tilde{W} = e^{-\hat{\mathcal{L}} y- \hat{\mathcal{Z}} t}(\tilde{V}_1 dx + \tilde{V}_2 dt) \tilde{\Phi}.
\end{equation}  
For each $n = 1, \dots, 18$, we define a solution $\tilde{\Phi}_n(x,t,k)$ of (\ref{tildePhilax}) by the following system of integral equations: 
\begin{equation}\label{tildePhinintegraleq}
(\tilde{\Phi}_n)_{ij}(x,t,k) = \delta_{ij} + \int_{\gamma_{ij}^n} \left(e^{\hat{\mathcal{L}} y(x,t) + \hat{\mathcal{Z}} t} \tilde{W}_n(x',t',k)\right)_{ij}, \qquad k \in D_n, \quad i,j = 1, 2,3,
\end{equation}
where $\tilde{W}_n$ is given by (\ref{tildeWdef}) with $\tilde{\Phi}$ replaced by $\tilde{\Phi}_n$.

%Recall that $\varkappa_j = e^{\frac{\pi i(j-1)}{3}}$, $j = 1,\dots, 6$ denote the six roots of unity and $K_j = e^{\frac{\pi i j}{3} - \frac{\pi i}{6}}$, $j = 1, \dots, 6$ denote the six points where $\lambda = 0$. 

\begin{proposition}\label{tildePhiprop}
For each $n = 1, \dots, 18$, the function $\tilde{\Phi}_n(x,t,k)$ is well-defined by equation (\ref{tildePhinintegraleq}) for $k \in D_n$ and $(x,t)$ in the domain (\ref{halflinedomain}). For any fixed point $(x,t)$, $\tilde{\Phi}_n$ is analytic as a function of $k \in D_n$ away from the set $\{\varkappa_j, K_j\}_1^6 \cup \{k_j\}$, where $\{k_j\}$ denotes a possibly empty discrete set of singularities at which the Fredholm determinant vanishes. 

Moreover, for $n = 1, \dots, 6$, the function $\tilde{\Phi}_n(x,t,k)$ is bounded for $k \in D_n$ away from the set $\{\varkappa_j, K_j\}_1^6 \cup \{k_j\}$ and it admits a bounded and continuous extension to $\bar{D}_n$ away from these points.
\end{proposition}
\proofbegin
The definition of $\{\gamma_j\}_1^3$ together with the assumption (\ref{g0assumption}) imply the following relations on the contours:
\begin{align}\nonumber 
& \gamma_1: \hspace{3.6cm} \qquad t - t' \leq 0,
	\\ \label{contourrelations2} 
& \gamma_2: y(x,t) - y(x',t') \geq 0, \qquad t - t' \geq 0,
	\\ \nonumber
& \gamma_3: y(x,t) - y(x',t') \leq 0, \qquad t - t' = 0,
\end{align}
The $(ij)$th entry of the integral equation (\ref{tildePhinintegraleq}) involves the exponential factor 
\begin{equation}\nonumber
e^{(l_i - l_j)(y(x,t) - y(x',t')) + (z_i - z_j)(t - t')}.
\end{equation}
The proof is now similar to that of Proposition \ref{Phiprop}, except that since $y(x,t) - y(x',t')$ can take on both signs in the case of $\gamma_1$, the exponential is not necessarily bounded for the integration along $\gamma_1$. However, the matrices $(\gamma^n)_{ij} := \gamma_{ij}^{n}$ for $n = 1, \dots, 6$, are given by
\begin{align*}
& \gamma^1 = \begin{pmatrix} 
\gamma_3 & \gamma_2 & \gamma_2 \\
\gamma_3 & \gamma_3 & \gamma_2 \\
\gamma_3 & \gamma_3 & \gamma_3 \end{pmatrix}, \qquad
\gamma^2 = \begin{pmatrix} 
 \gamma_3 & \gamma_2 & \gamma_2 \\
 \gamma_3 & \gamma_3 & \gamma_3 \\
 \gamma_3 & \gamma_2 & \gamma_3
\end{pmatrix}, \qquad
\gamma^3 = \begin{pmatrix} 
 \gamma_3 & \gamma_2 & \gamma_3 \\
 \gamma_3 & \gamma_3 & \gamma_3 \\
 \gamma_2 & \gamma_2 & \gamma_3
\end{pmatrix},
	\\
& \gamma^4 = \begin{pmatrix} 
 \gamma_3 & \gamma_3 & \gamma_3 \\
 \gamma_2 & \gamma_3 & \gamma_3 \\
 \gamma_2 & \gamma_2 & \gamma_3
\end{pmatrix}, \qquad
\gamma^5 = \begin{pmatrix} 
 \gamma_3 & \gamma_3 & \gamma_3 \\
 \gamma_2 & \gamma_3 & \gamma_2 \\
 \gamma_2 & \gamma_3 & \gamma_3
\end{pmatrix}, \qquad
\gamma^6 = \begin{pmatrix} 
 \gamma_3 & \gamma_3 & \gamma_2 \\
 \gamma_2 & \gamma_3 & \gamma_2 \\
 \gamma_3 & \gamma_3 & \gamma_3
\end{pmatrix}.
\end{align*}
Therefore, the definition of $\tilde{\Phi}_n$ for $k \in D_n$, $n = 1,\dots, 6$, does not involve integration along $\gamma_1$, so we can still conclude that $\tilde{\Phi}_n$ has the stated boundedness properties for $n = 1, \dots, 6$. 
\proofend

The proof of the following lemma is similar to that of Lemma \ref{Phinsymmlemma}.

\begin{lemma}\label{tildePhinsymmlemma} 
Define the sectionally meromorphic function $\tilde{\Phi}_*(x,t,k)$ by
$$\tilde{\Phi}_*(x,t,k) = \tilde{\Phi}_n(x,t,k), \qquad k \in D_n.$$
Then
\begin{align*}
&  \tilde{\Phi}_*(k) = \mathcal{A} \tilde{\Phi}_*(\omega k)\mathcal{A}^{-1}, \quad  \tilde{\Phi}_*(k) = \mathcal{B} \overline{\tilde{\Phi}_*(\overline{k})}\mathcal{B},	\quad
\tilde{\Phi}_*(k) = \mathcal{B} \tilde{\Phi}_*(1/k)\mathcal{B}, \quad k \in \C.
\end{align*}
\end{lemma}

According to Proposition \ref{tildePhiprop}, the $\tilde{\Phi}_n$'s are bounded near $k = \infty$ and $k=0$. The following lemma determines the leading behavior of the $\tilde{\Phi}_n$'s near these points.

\begin{lemma}
The eigenfunctions $\tilde{\Phi}_n(x,t,k)$, $n = 1, \dots, 6$, satisfy
\begin{align}\label{tildePhinatinfty}
&  \tilde{\Phi}_n(x,t,k) = I + O(1/k) \quad \text{as} \quad k \to \infty,
 	\\\label{tildePhinat0}
&  \tilde{\Phi}_n(x,t,k) = I + O(k) \quad \text{as} \quad k \to 0.
\end{align}
\end{lemma}
\proofbegin
We substitute the expansion
$$\tilde{\Phi}(x,t,k) = \tilde{\Phi}^{(0)}(x,t) + \tilde{\Phi}^{(1)}(x,t) k + \dots, \qquad k \to 0,$$
where the $\tilde{\Phi}^{(j)}$'s are independent of $k$, into the Lax pair (\ref{tildePhilax}). 
In view of Lemma \ref{tildeV1V2lemma} and the fact that 
$$\mathcal{L} = \frac{1}{\sqrt{3}} \begin{pmatrix} \omega^2 & 0 & 0 \\ 0 & \omega & 0 \\ 0 & 0 & 1 \end{pmatrix}\frac{1}{k} + O(k), \qquad k \to 0,$$
the terms of $O(1/k)$ of the $x$-part imply that $\tilde{\Phi}^{(0)}(x,t)$ is a diagonal matrix. 
The diagonal terms of $O(1)$ of the $x$ and $t$-parts then imply that $\tilde{\Phi}^{(0)}(x,t)$ is independent of $(x,t)$. Evaluation at the points $(x,t) = (x_j, t_j)$, $j = 1,2,3$, yields $\tilde{\Phi}^{(0)}(x,t) = I$. This proves (\ref{tildePhinat0}). Equation (\ref{tildePhinatinfty}) follows by symmetry.
\proofend

\begin{figure}
\begin{center}
 \begin{overpic}[width=.85\textwidth]{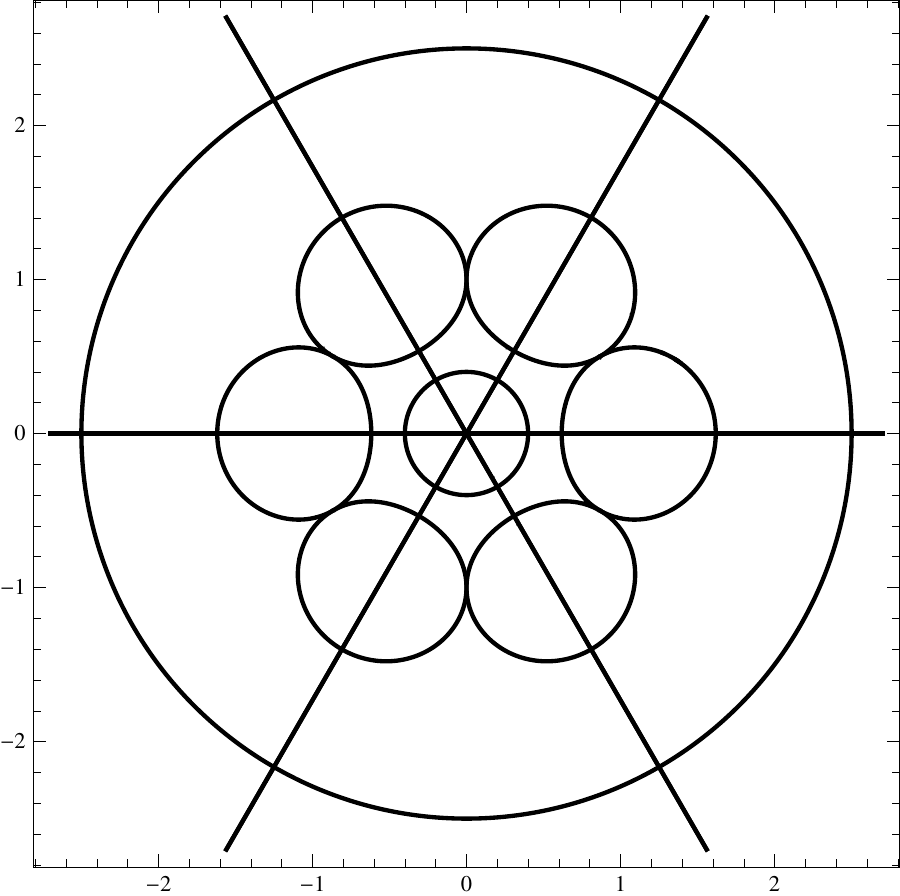}
      \put(80,66){\footnotesize $E_1$}
      \put(59,55.5){\footnotesize $E_1$}
      \put(50,84){\footnotesize $E_2$}
      \put(50,60){\footnotesize $E_2$}
     \put(22,67){\footnotesize $E_3$}
      \put(42,55.5){\footnotesize $E_3$}
     \put(22,34){\footnotesize $E_4$}
      \put(42,46){\footnotesize $E_4$}
      \put(50,18){\footnotesize $E_5$}
      \put(50,41){\footnotesize $E_5$}
      \put(80,34){\footnotesize $E_6$}
      \put(59,46){\footnotesize $E_6$}
      \put(68,55){\footnotesize $E_7$}
      \put(64,65){\footnotesize $E_8$}
      \put(56,69){\footnotesize $E_9$}
      \put(44,69){\footnotesize $E_{10}$}
      \put(37,64){\footnotesize $E_{11}$}
      \put(31,55){\footnotesize $E_{12}$}
      \put(31,46){\footnotesize $E_{13}$}
      \put(37,37){\footnotesize $E_{14}$}
      \put(44,32){\footnotesize $E_{15}$}
      \put(56,32){\footnotesize $E_{16}$}
      \put(63,36){\footnotesize $E_{17}$}
      \put(68,46){\footnotesize $E_{18}$}
      \put(54,53){\tiny $E_{19}$}
      \put(50,55.5){\tiny $E_{20}$}
      \put(46,53){\tiny $E_{21}$}
      \put(46,49){\tiny $E_{22}$}
      \put(50,46){\tiny $E_{23}$}
      \put(54,49){\tiny $E_{24}$}
      \put(90,85){\footnotesize $E_{19}$}
      \put(50,95.5){\footnotesize $E_{20}$}
      \put(10,85){\footnotesize $E_{21}$}
      \put(10,15){\footnotesize $E_{22}$}
      \put(50,5.5){\footnotesize $E_{23}$}
      \put(90,15){\footnotesize $E_{24}$}
 \end{overpic}
    \qquad \qquad
     \begin{figuretext}\label{Ens.pdf}
       The sets $E_n$, $n = 1, \dots, 24$, which decompose the complex $k$-plane. 
         \end{figuretext}
     \end{center}
\end{figure}

%\begin{remark}\upshape
%As is customary in the implementation of the IST, we will assume that the discrete set of singularities $\{k_j\} \subset D_n$ of $\Psi_n$ consists of a finite number of simple poles in the interior of $D_n$. 
%\end{remark}

\section{A sectionally meromorphic function}\label{specsec}\nequation
Let $\{\Phi_n\}_1^{18}$ and $\{\tilde{\Phi}_n\}_{1}^{18}$ denote the eigenfunctions defined in Section \ref{eigensec}. We have good control over the $\Phi_n$'s near $K_j = e^{\frac{\pi i j}{3} - \frac{\pi i}{6}}$, $j = 1, \dots, 6$, but not near $k = \infty$ and $k = 0$. On the other hand, we have good control over the $\tilde{\Phi}_n$'s near $k=\infty$ and $k = 0$, but not near the $K_j$'s. Therefore, we will introduce a radius $R > 2$ and formulate a RH problem by using the $\Phi_n$'s for $1/R < |k| < R$ and the $\tilde{\Phi}_n$'s for $|k| < 1/R$ and for $|k| > R$. 

Let $R>2$. Define sets $\{E_n\}_1^{24}$ by (see Figure \ref{Ens.pdf})
\begin{align}\nonumber
& E_n = D_n \cap \{1/R < |k| < R\}, \qquad n = 1, \dots, 18,
	\\ \label{Endef}
& E_{n+18} = D_{n} \cap \{|k| < 1/R \text{  or  } |k|> R\}, \qquad n = 1, \dots, 6.
\end{align}	

Since the map $F:(x,t) \mapsto (y,t)$, $y = y(x,t)$, is a bijection from the domain $\Omega = \{0 \leq x < \infty, 0 \leq t < T\}$ onto $F(\Omega)$, we can define functions $\{M_n(y, t, k)\}_1^{24}$ for $(y,t) \in F(\Omega)$ by
\begin{align}\label{Mndef}
M_n(y, t, k) = \begin{cases} \Phi_n(x, t, k)e^{\mathcal{L}(k)(x - y)}, \qquad k \in E_n, \quad n = 1, \dots, 18, \\
P^{-1}(k)D(x,t) P(k) \tilde{\Phi}_{n-18}(x, t, k), \qquad k \in E_n, \quad n = 19, \dots, 24.
 \end{cases}
\end{align}
The $M_n$'s defined in (\ref{Mndef}) are bounded and analytic everywhere on the Riemann $k$-sphere, except at the six roots of unity $\{\varkappa_j\}_1^6$ and at the $k_j$'s (note that the combination $P^{-1}(k)D(x,t) P(k)$ is analytic at $k = \infty$ and $k = 0$). We deal with the singularities at the $\varkappa_j$'s by formulating the RH problem in terms of the row vectors $\nu_n$ defined by (see \cite{BS2011})
\begin{equation}\label{nundef}
\nu_n(y, t, k) = \begin{pmatrix} 1 & 1 & 1 \end{pmatrix} M_n(y,t,k), \qquad k \in E_n, \quad n = 1, \dots, 24.
\end{equation}
Lemma \ref{polestructurelemma} implies that the $\nu_n$'s are bounded near the $\varkappa_j$'s.
In Section \ref{residuesec}, we deal with the singularities at the $\varkappa_j$'s by deriving appropriate residue conditions.

Let $M$ and $\nu$ denote the sectionally meromorphic functions on the Riemann $k$-sphere which equal $M_n$ and $\nu_n$ respectively for $k \in E_n$.

\begin{lemma}\label{Msymmlemma}
The function $M$ obeys the symmetries
\begin{align*}
&  M(k) = \mathcal{A} M(\omega k)\mathcal{A}^{-1}, \quad  M(k) = \mathcal{B} \overline{M(\overline{k})}\mathcal{B},	\quad
M(k) = \mathcal{B} M(1/k)\mathcal{B}, \quad k \in \C,
\end{align*}
where $\mathcal{A}$, $\mathcal{B}$ are defined in (\ref{ABdef}) and we have suppressed the $(x,t)$-dependence.
\end{lemma}
\proofbegin
This is a consequence of equation (\ref{Psymmetries}) and the symmetry properties of the $\Phi_n$'s and the $\tilde{\Phi}_n$'s.
\proofend

We define spectral functions $S_n(k)$ by
\begin{align}\label{Sndef}  
  & S_n(k) = M_n(0, 0,k), \qquad k \in E_n, \quad n = 1, \dots, 24.
\end{align}
The tracelessness of the matrices $\{V_j, \tilde{V}_j\}_1^2$ implies that $\det \Phi_n \equiv 1$ and $\det \tilde{\Phi}_n \equiv 1$. In particular,  
$$\det S_n(k) = 1, \qquad n = 1, \dots, 24.$$
The exponential factor $e^{\mathcal{L}(x - y)}$ on the right-hand side of (\ref{Mndef}) has been included because it ensures that the jump matrices introduced in the next proposition depend on $x$ only through the function $y(x,t)$.

\begin{proposition}
For each $n = 1, \dots, 24$, the function $\nu_n$  is bounded and analytic in $E_n$ away from the possibly empty discrete set $\{k_j\}$. Moreover, each $\nu_n$ has a continuous and bounded extension to $\bar{E}_n$. 
The function $\nu$ satisfies the jump conditions
\begin{equation}\label{MnMmrelation}  
  \nu_n = \nu_m J_{m,n}, \qquad k \in \bar{E}_n \cap \bar{E}_m, \qquad n, m = 1, \dots, 24, \quad n \neq m, 
\end{equation}
where the jump matrix $J_{m,n}(y, t, k)$, $J_{m,n} = J_{n,m}^{-1}$, is defined by
\begin{equation}\label{Jmndef}
J_{m,n} = e^{\hat{\mathcal{L}} y + \hat{\mathcal{Z}} t}(S_m^{-1}S_n),\qquad n, m \in \{1, \dots, 24\}.
\end{equation}
\end{proposition}
\proofbegin
The analyticity and boundedness properties of the $\nu_n$'s follow from the properties of the $\Phi_n$'s and the $\tilde{\Phi}_n$'s established in Section \ref{eigensec}.

Suppose $n, m \in \{1, \dots, 24\}$. Equations (\ref{Phidef}), (\ref{tildePhidef}), and (\ref{Mndef}) imply that the functions $P M_n e^{\mathcal{L} y + \mathcal{Z} t}$ and $P M_m e^{\mathcal{L} y + \mathcal{Z} t}$ both satisfy the Lax pair equations (\ref{psilax}). Thus, there exists a matrix $J(k)$ independent of $x, t$ such that
\begin{align}\label{MnMm}
M_n e^{\mathcal{L} y + \mathcal{Z} t} = M_m e^{\mathcal{L} y + \mathcal{Z} t} J(k).
\end{align}
Evaluation at $x = t = 0$ yields $J = S_m^{-1}S_n$. Multiplying (\ref{MnMm}) by $(1,1,1)$ from the left, we obtain the jump condition (\ref{MnMmrelation}) with $J_{m,n}$ given by (\ref{Jmndef}). 
\proofend

\section{Residue conditions}\label{residuesec}\nequation
If the $\nu_n$'s have pole singularities at some points $\{k_j\}$, $k_j \in \C$, the RH problem needs to include the residue conditions at these points. 
We will assume that all $k_j$'s lie in the interiors of the sets $\{E_n\}_1^{18}$; singularities in the interiors of the sets $\{E_n\}_{19}^{24}$ can be avoided by choosing $R>0$ large enough or can be treated by a similar argument.
The residue conditions can be found by relating the $M_n$'s to another set of solutions of (\ref{Philax}), denoted by $\{\mu_j\}_1^3$, which are defined by 
\begin{equation}\label{mujdef}
  \mu_j(x,t,k) = I +  \int_{\gamma_j} e^{\hat{\mathcal{L}} x + \hat{\mathcal{Z}} t} W_j(x',t',k), \qquad j = 1, 2,3,
\end{equation}
where $\{\gamma_j\}_1^3$ are the contours shown in FigureÊ \ref{mucontours.pdf} and $W_j$ is given by (\ref{Wdef}) with $\Phi$ replaced with $\mu_j$.
The functions $\mu_1$ and $\mu_2$ are defined for all $k$, whereas $\mu_3$, whose definition involves integration from Ê$x = \infty$, is defined for $k \in (\mathcal{S}, \omega^2 \mathcal{S}, \omega \mathcal{S})$, where
$$\mathcal{S} = \{k | 0 \leq \arg(k) \leq 2\pi/3\}.$$
Here the notation $k \in (\mathcal{S}, \omega^2 \mathcal{S}, \omega\mathcal{S})$ indicates that the first, second, and third columns are valid for $k$ in the sets $\mathcal{S}$, $\omega^2 \mathcal{S}$, and $\omega\mathcal{S}$, respectively. 
 
Since (\ref{mujdef}) are Volterra integral equations, the analyticity properties of $W$ imply that away from the points $\{\varkappa_j\}_1^6$, $\mu_1$ and $\mu_2$ are analytic functions of $k \in E_n$ with continuous and bounded extensions to $\bar{E}_n$, $n = 1, \dots, 18$. The argument in the proof of Lemma \ref{Phinasymptoticslemma} implies that $\mu_1$ and $\mu_2$ tend to the identity matrix as $k \to K_j$. Analogous statements apply to the column vectors of $\mu_3$ within their respective domains of definition.

\subsection{A matrix factorization problem}
We introduce spectral functions $s(k)$ and $S(k)$ by
\begin{equation}\label{mu3mu2mu1sS}
\mu_3 = \mu_2 e^{\hat{\mathcal{L}} x + \hat{\mathcal{Z}} t}s(k), \qquad
\mu_1 = \mu_2 e^{\hat{\mathcal{L}} x + \hat{\mathcal{Z}} t}S(k),
\end{equation}
that is,
\begin{equation}\label{sSdef}  
  s(k) = \mu_3(0,0,k), \qquad S(k) = \mu_1(0,0,k).
\end{equation}
Lemma \ref{symmlemma} together with the initial conditions
$$\mu_j(x_j, t_j, k) = I, \qquad j = 1,2,3,$$
imply that the eigenfunctions $\mu_j$, $j = 1,2,3$, and hence also $s(k)$ and $S(k)$, obey the symmetries of equation (\ref{Phisymmetries}). 
Defining spectral functions $R_n$ and $T_n$, $n = 1, \dots, 18$, by
\begin{subequations}\label{Sn13def}  
\begin{align}
 & R_n(k) = e^{- \hat{\mathcal{Z}} T}M_n(y(0,T), T,k)e^{\mathcal{L} y(0,T)}, \qquad k \in E_n, 
 	\\
  &T_n(k) = \lim_{x \to \infty} e^{-\mathcal{L}x}M_n(y, 0,k)e^{\mathcal{L}y}, \qquad k \in E_n,
\end{align}
\end{subequations}
we have
\begin{align}\label{Mnmujrelations}
\begin{cases}
  M_n(y,t,k) =\mu_1(x,t,k) e^{\mathcal{L}x + \mathcal{Z} t} R_n(k) e^{-\mathcal{L}y - \mathcal{Z} t}, 	\\  M_n(y,t,k) =\mu_2(x,t,k) e^{\mathcal{L}x + \mathcal{Z} t} S_n(k) e^{-\mathcal{L}y - \mathcal{Z} t}, 	\\
M_n(y,t,k) =\mu_3(x,t,k) e^{\mathcal{L}x + \mathcal{Z} t} T_n(k) e^{-\mathcal{L}y - \mathcal{Z} t}, 
  \end{cases}
  \quad j = 1, 2,3, \quad n = 1, \dots, 18,
\end{align}  
and
\begin{align} \nonumber
& \left(R_n(k)\right)_{ij} = 0 \quad \text{if} \quad \gamma_{ij}^n = \gamma_1,
	\\\label{Sn123ij0}
& \left(S_n(k)\right)_{ij} = 0 \quad \text{if} \quad \gamma_{ij}^n = \gamma_2,
	\\ \nonumber
& \left(T_n(k)\right)_{ij} = \delta_{ij} \quad \text{if} \quad \gamma_{ij}^n = \gamma_3,
\end{align}
where $\gamma_{ij}^n$ is defined by (\ref{gammaijnudef}). 
We can now find expressions for the $S_n$'s in terms of the entries of $s$ and $S$ by solving the matrix factorization problem\footnote{Strictly speaking, the first equation in (\ref{sSSnrelations}) is defined only for $k \in (\mathcal{S}, \omega^2 \mathcal{S}, \omega \mathcal{S})$; however, this problem can be circumvented by introducing a new spectral function $s(k; X_0)$ defined for all $k \in \C$ and then letting $X_0 \to \infty$, see \cite{L2012}.}
\begin{align}\label{sSSnrelations}  
  s(k) = S_n(k)T_n^{-1}(k), \qquad S(k) = S_n(k)R_n^{-1}(k), \qquad k \in E_n.
\end{align}
The conditions (\ref{Sn123ij0}) imply that (\ref{sSSnrelations}) are $18$ scalar equations for $18$ unknowns. Solving this system of equations, we find the following result.

\begin{proposition}\label{Snexplicitlemma}
 The spectral functions $S_1$, $S_7$, and $S_8$ defined by (\ref{Sndef}) can be expressed in terms of the entries of $s(k)$ and $S(k)$ as follows:
 \begin{subequations}\label{Snexplicit}
\begin{align}\label{S1explicit}
&  S_1(k) = \begin{pmatrix}
s_{11} & 0 & 0 \\
s_{21} & \frac{m_{33}(s)}{s_{11}} & 0 \\
s_{31} & \frac{m_{23}(s)}{s_{11}} & \frac{1}{m_{33}(s)} 
  \end{pmatrix},
	\\ \label{S7explicit}
&  S_7(k) =  \begin{pmatrix}
s_{11} & \frac{m_{33}(s) m_{21}(S)-m_{23}(s)m_{31}(S)}{W_1} & 0 \\
s_{21} & \frac{m_{33}(s) m_{11}(S)-m_{13}(s)m_{31}(S)}{W_1}  & 0 \\
s_{31} & \frac{m_{23}(s) m_{11}(S)-m_{13}(s)m_{21}(S)}{W_1}  & \frac{1}{m_{33}(s)}
\end{pmatrix},
  	\\ \label{S8explicit}
&  S_8(k) =  \begin{pmatrix}
s_{11} & 0 & 0 \\
s_{21} & \frac{m_{33}(s)}{s_{11}} & \frac{m_{32}(S)}{W_2} \\
s_{31} & \frac{m_{23}(s)}{s_{11}} & \frac{m_{22}(S)}{W_2}
\end{pmatrix},
\end{align}
\end{subequations}
where the functions $\{W_j(k)\}_1^2$ are defined by
\begin{align*}
& W_1(k) = s_{11} m_{11}(S) - s_{21}m_{21}(S) + s_{31}m_{31}(S),
	\\
& W_2(k) = m_{33}(s)m_{22}(S) - m_{23}(s)m_{32}(S),
\end{align*}
$m_{ij}(s)$ and $m_{ij}(S)$ denote the $(ij)$th minors of the matrices $s(k)$ and $S(k)$ respectively (i.e. $m_{ij}(s)$ equals the determinant of the $2 \times 2$-matrix obtained from $s$ by deleting the $i$th row and the $j$th column), and the $k$-dependence has been suppressed for clarity. 
\end{proposition}

The spectral function $S_n(k)$ for any $n = 1, \dots, 18$, can be obtained from (\ref{Snexplicit}) together with the symmetries
\begin{align}\label{Snsymm}
&  S_*(k) = \mathcal{A} S_*(\omega k)\mathcal{A}^{-1}, \quad  S_*(k) = \mathcal{B} \overline{S_*(\overline{k})}\mathcal{B},	\quad
S_*(k) = \mathcal{B} S_*(1/k)\mathcal{B},
\end{align}
where $S_*$ denotes the sectionally meromorphic function defined by
$$S_*(x,t,k) = S_n(x,t,k), \qquad k \in E_n.$$

\begin{remark}\upshape
Although $s(k)$ is only defined by (\ref{mujdef}) for $k \in (\mathcal{S}, \omega^2 \mathcal{S}, \omega \mathcal{S})$, the functions $\{m_{j3}(s(k))\}_{j=1}^3$ in (\ref{Snexplicit}) can be extended by analytic continuation and are well-defined for $k \in E_1 \cup E_7 \cup E_8$. 
Indeed, if $\mu$ satisfies the Lax pair (\ref{Philax}), then the cofactor eigenfunction $\mu^A$ defined by 
$$\mu^{A} = \begin{pmatrix} m_{11}(\mu) & -m_{12}(\mu) & m_{13}(\mu) \\
-m_{21}(\mu) & m_{22}(\mu) & -m_{23}(\mu) \\
m_{31}(\mu) & -m_{32}(\mu) & m_{33}(\mu)
\end{pmatrix}$$
satisfies the Lax pair
\begin{align}\label{mmulax}
\begin{cases}
\mu^A_x + [\mathcal{L}, \mu^A] = -V_1^T\mu^A, \\
\mu^A_t + [\mathcal{Z}, \mu^A] = -V_2^T\mu^A.
\end{cases}
\end{align}
Thus, the eigenfunctions $\{\mu_j^A\}_1^3$ satisfy the Volterra integral equations
\begin{equation}\label{mujAdef}
  \mu_j^A(x,t,k) = I -  \int_{\gamma_j} e^{-\hat{\mathcal{L}} (x-x') - \hat{\mathcal{Z}}(t-t')} (V_1^T dx + V_2^T dt)\mu_j^A, \qquad j = 1, 2,3.
\end{equation}
The third column of $\mu_3^A$ can be defined by (\ref{mujAdef}) for $k$ such that $l_3(k)$ is larger than or equal to both $l_1(k)$ and $l_2(k)$. This set includes all $k \in E_1 \cup E_7 \cup E_8$. 
\end{remark}

\subsection{The residue conditions}
It follows from (\ref{Mnmujrelations}) and the analyticity properties of $\mu_2$ that $\nu$ can only have singularities at the points $\{k_j\}$ where the $S_n$'s have singularities. 
In view of the symmetries of Lemma \ref{symmlemma}, it is enough to study the case of $k_j \in E_1 \cup E_7 \cup E_8$. We infer from the explicit formulas (\ref{Snexplicit}) that the possible singularities of $M$ in $E_1 \cup E_7 \cup E_8$ are as follows:
\begin{itemize}
\item $[M]_2$ could have poles in $E_1 \cup E_8$ at the zeros of $s_{11}(k)$.
\item $[M]_3$ could have poles in $E_1 \cup E_7$ at the zeros of $m_{33}(s(k))$.
\item $[M]_2$ could have poles in $E_7$ at the zeros of $W_1(k)$.
\item $[M]_3$ could have poles in $E_8$ at the zeros of $W_2(k)$.
\end{itemize}
We denote the above possible zeros by $\{k_j\}_1^N$ and assume they satisfy the following assumption.

\begin{assumption}\label{kjass}\upshape
We assume that
\begin{itemize}
\item $s_{11}(k)$ has $n_1$ simple zeros in $E_1 \cup E_8$ denoted by $\{k_j\}_1^{n_1}$,
\item $m_{33}(s(k))$ has $n_2 - n_1$ simple zeros in $E_1$ denoted by $\{k_j\}_{n_1 + 1}^{n_2}$,
\item $m_{33}(s(k))$ has $n_3 - n_2$ simple zeros in $E_7$ denoted by $\{k_j\}_{n_2 + 1}^{n_3}$,
\item $W_1(k)$ has $n_4 - n_3$ simple zeros in $E_7$ denoted by $\{k_j\}_{n_3 + 1}^{n_4}$,
\item $W_2(k)$ has $N - n_4$ simple zeros in $E_8$ denoted by $\{k_j\}_{n_4 + 1}^{N}$,
\end{itemize}
and that none of these zeros coincide. Moreover, we assume that none of these functions have zeros on the boundaries of the $E_n$'s.  
We also assume, for simplicity, that $R > 0$ has been chosen so large in (\ref{Endef}) that there are no pole singularities in $E_n$, $n = 19, \dots, 24$.
\end{assumption}

In the next proposition we determine the residue conditions at these zeros.

\begin{proposition}
Let $M$ be the sectionally meromorphic function defined by (\ref{Mndef}) and assume that the set $\{k_j\}_1^N$ of singularities in $E_1 \cup E_7 \cup E_8$ are as in assumption \ref{kjass}. Then the following residue conditions hold: 
\begin{subequations}\label{Mres}
\begin{align} \label{Mresa}
\underset{k_j}{\res} [M]_2 = &\; \frac{m_{33}(s(k_j))}{\dot{s}_{11}(k_j) s_{21}(k_j)} e^{\theta_{12}(k_j)} [M(k_j)]_1, \qquad 1 \leq j \leq n_1, \; k_j \in E_1 \cup E_8,
	\\ \label{Mresb}
\underset{k_j}{\res} [M]_3 = &\; \frac{s_{11}(k_j) e^{\theta_{23}(k_j)} }{\dot{m}_{33}(s(k_j)) m_{23}(s(k_j)) } [M(k_j)]_2, \qquad n_1 < j \leq n_2, \; k_j \in E_1,
	\\ \nonumber
\underset{k_j}{\res} [M]_3 = &\; \frac{1}{\dot{m}_{33}(s(k_j))}\biggl(
\frac{m_{31}(S(k_j))e^{\theta_{13}(k_j)}}{W_1(k_j)}  [M(k_j)]_1
+ \frac{s_{11}(k_j)e^{\theta_{23}(k_j)} }{m_{23}(s(k_j))} [M(k_j)]_2\biggr), 
	\\ \label{Mresc}
& \hspace{5cm} n_2 < j \leq n_3, \; k_j \in E_7,
	\\ \nonumber
\underset{k_j}{\res} [M]_2 = &\; \frac{m_{33}(s(k_j))m_{21}(S(k_j)) - m_{23}(s(k_j))m_{31}(S(k_j))}{\dot{W}_1(k_j) s_{11}(k_j)} e^{\theta_{12}(k_j)} [M(k_j)]_1, 
	\\ \label{Mresd}
& \hspace{5cm} n_3 < j \leq n_4,\; k_j \in E_7,
	\\ \label{Mrese}
\underset{k_j}{\res} [M]_3 = &\; \frac{m_{32}(S(k_j)) s_{11}(k_j) e^{\theta_{23}(k_j)} }{\dot{W}_2(k_j) m_{33}(s(k_j))} [M(k_j)]_2, \qquad n_4 < j \leq N, \; k_j \in E_8,
\end{align}
\end{subequations}
where $\dot{f} := df/dk$ and
$$\theta_{ij}(k) = (l_i(k) - l_j(k))y + (z_i(k) - z_j(k))t.$$
\end{proposition}
\proofbegin
We will prove (\ref{Mresa}) and (\ref{Mrese}); the conditions (\ref{Mresb})-(\ref{Mresd}) follow by similar arguments.
Equation (\ref{Mnmujrelations}) implies the relations
\begin{equation}\label{M1mu2rel}  
  M_1 = \mu_2 e^{\mathcal{L} x + \mathcal{Z} t} S_1 e^{-\mathcal{L}y - \mathcal{Z} t} \quad
 \text{and} \quad  M_8 = \mu_2 e^{\mathcal{L} x + \mathcal{Z} t} S_8 e^{-\mathcal{L}y - \mathcal{Z} t}.
\end{equation}
For $i,j= 1,2,3$, let $\tilde{\theta}_{ij} = (l_i - l_j)x + (z_i - z_j)t.$
In view of the expressions for $S_1$ and $S_8$ given in (\ref{Snexplicit}), the three columns of (\ref{M1mu2rel}a) read
\begin{subequations}
\begin{align}\label{M1mu2a}
& [M_1]_1e^{l_1(y - x)} = s_{11} [\mu_2]_1  
+ s_{21} e^{\tilde{\theta}_{21}} [\mu_2]_2
+ s_{31} e^{\tilde{\theta}_{31}} [\mu_2]_3 ,
	\\ \label{M1mu2b}
 & [M_1]_2e^{l_2(y - x)} = \frac{m_{33}(s)}{s_{11}} [\mu_2]_2 
  + \frac{m_{23}(s)}{s_{11}} e^{\tilde{\theta}_{32}} [\mu_2]_3, 
   	\\ \label{M1mu2c}
& [M_1]_3 e^{l_3(y - x)} = \frac{1}{m_{33}(s)} [\mu_2]_3,
\end{align}
\end{subequations} 
while the three columns of (\ref{M1mu2rel}b) read
 \begin{subequations}
\begin{align}\label{M8mu2a}
& [M_8]_1e^{l_1(y - x)} = s_{11} [\mu_2]_1  
+ s_{21} e^{\tilde{\theta}_{21}} [\mu_2]_2 
+ s_{31} e^{\tilde{\theta}_{31}} [\mu_2]_3,
	\\ \label{M8mu2b}
 & [M_8]_2e^{l_2(y - x)} = \frac{m_{33}(s)}{s_{11}} [\mu_2]_2 
  + \frac{m_{23}(s)}{s_{11}} e^{\tilde{\theta}_{32}} [\mu_2]_3, 
   	\\ \label{M8mu2c}
& [M_8]_3 e^{l_3(y - x)}  = \frac{m_{32}(S)}{W_2} e^{\tilde{\theta}_{23}} [\mu_2]_2
+ \frac{m_{22}(S)}{W_2} [\mu_2]_3.
\end{align}
\end{subequations}  

In order to prove (\ref{Mresa}), we first suppose thatÊ $k_j \in E_1$ is a simple zero of $s_{11}(k)$.
Solving (\ref{M1mu2a}) and (\ref{M1mu2c}) for $[\mu_2]_2$ and $[\mu_2]_3$ and substituting the result into (\ref{M1mu2b}), we find
$$ [M_1]_2 = \frac{m_{33}(s)}{s_{11}s_{21}}e^{\theta_{12}}[M_1]_1
- \frac{m_{33}(s)}{s_{21}}e^{\tilde{\theta}_{12} + l_2(x-y)}[\mu_2]_1
+ \frac{m_{13}(s)m_{33}(s)}{s_{21}} e^{\theta_{32}} [M_1]_3.$$
Taking the residue of this equation at $k_j$, we find the condition (\ref{Mresa}) in the case when $k_j \in E_1$. 
Similarly, solving (\ref{M8mu2a}) and (\ref{M8mu2c}) for $[\mu_2]_2$ and $[\mu_2]_3$, substituting the result into (\ref{M8mu2b}), and taking the residue at $k_j$, we find after long computations that (\ref{Mresa}) holds also if $k_j$ is a simple zero of $s_{11}$ in $E_8$.

In order to prove (\ref{Mrese}), we suppose thatÊ $k_j \in E_8$ is a simple zero of $W_2(k)$. 
Solving (\ref{M8mu2a}) and (\ref{M8mu2b}) for $[\mu_2]_2$ and $[\mu_2]_3$, substituting the result into (\ref{M8mu2c}), and taking the residue at $k_j$, we find (\ref{Mrese}).
\proofend

\section{The Riemann-Hilbert problem}\label{RHsec}\nequation
The sectionally meromorphic function $\nu(y,t,k)$ defined in Section \ref{specsec} satisfies a Riemann-Hilbert problem which can be formulated in terms of the initial and boundary values of $u(x,t)$. By solving this RH problem, the solution $u(x,t)$ of (\ref{DP}) in the half-line domain (\ref{halflinedomain}) can be recovered in parametric form.

\begin{theorem}\label{RHtheorem}
Suppose that $u(x,t)$ is a solution of (\ref{DP}) in the half-line domain $\{0 \leq x < \infty, 0 \leq t < T\}$ with sufficient smoothness and decay as $x \to \infty$. Suppose that the initial and boundary values $\{q_0(x), g_0(t), g_1(t), g_2(t)\}$ defined in (\ref{boundaryvalues}) satisfy the assumptions (\ref{qassumptions}) and (\ref{g0assumption}). Then $u(x,t)$ can be reconstructed from $\{q_0(x), g_0(t), g_1(t), g_2(t)\}$ as follows.

Use the initial and boundary data to define $\{\Phi_n(0,0, k)\}_1^{18}$ and $\{\tilde{\Phi}_n(0,0, k)\}_{1}^{6}$ via the integral equations (\ref{Phinintegraleq}) and (\ref{tildePhinintegraleq}), respectively. Define spectral functions $S_n$, $n = 1, \dots, 24$, by
\begin{align*}
& S_n(k) = \Phi_n(0,0,k), && k \in E_n, \quad n = 1, \dots, 18,
	\\
& S_n(k) = P^{-1}(k) D(0,0) P(k) \tilde{\Phi}_{n-18}(0,0,k), && k \in E_n, \quad n = 19, \dots, 24,
\end{align*}
where $P(k)$ and $D(x,t)$ are defined in (\ref{Pdef}) and (\ref{Ddef}) respectively.
Define the jump matrices $J_{m,n}(y, t, k)$, $n,m = 1, \dots, 24$, in terms of the $S_n$'s by equation (\ref{Jmndef}). 
Define the spectral functions $s(k)$ and $S(k)$ by equation (\ref{sSdef}). 
Assume that the possible zeros $\{k_j\}_1^N$ of the functions $s_{11}(k)$, $m_{33}(s(k))$, and $\{W_j(k)\}_1^2$ are as in assumption \ref{kjass}. 
 
Then the solution $u(x,t)$ is given in parametric form by
\begin{equation}\label{recoverq}
u(X(y,t),t) = -\frac{\partial}{\partial t} \log\bigl[\nu^1(y,t, e^{\frac{\pi i}{6}})\bigr], \qquad
X(y,t) = y - \log\bigl[\nu^1(y,t,e^{\frac{\pi i}{6}})\bigr],
\end{equation}
where the row-vector valued function $\nu = (\nu^1, \nu^2, \nu^3)$ satisfies the following RH problem:
\begin{itemize}
\item $\nu(y,t,k)$ is sectionally meromorphic on the Riemann $k$-sphere with jumps across the contours $\bar{E}_n \cap \bar{E}_m$, $n, m = 1, \dots, 24$, see Figure \ref{Ens.pdf}.

\item Across the contours $\bar{E}_n \cap \bar{E}_m$, $n, m = 1, \dots, 24$, $\nu$ satisfies the jump condition (\ref{MnMmrelation}).

\item $\nu$ satisfies the normalization condition:
$$\nu^j\bigl(y,t, e^{\frac{\pi i j}{3} - \frac{\pi i}{2}}\bigr) = 1, \qquad j = 1,2,3.$$

\item $\nu^2$ has simple poles at $k = k_j$ for $1 \leq j \leq n_1$ and $n_3 < j \leq n_4$. $\nu^3$ has simple poles at $k = k_j$ for $n_1  < j \leq n_3$ and $n_4 < j \leq N$. 
The associated residues satisfy the following residue conditions:
{\allowdisplaybreaks
\begin{subequations}\label{nures}
\begin{align} \label{nuresa}
\underset{k_j}{\res} \nu^2(y,t,k) = &\; \frac{m_{33}(s(k_j))}{\dot{s}_{11}(k_j) s_{21}(k_j)} e^{\theta_{12}(k_j)} \nu^1(y,t,k_j), \qquad 1 \leq j \leq n_1, \; k_j \in E_1 \cup E_8,
	\\ \label{nuresb}
\underset{k_j}{\res} \nu^3(y,t,k) = &\; \frac{s_{11}(k_j) e^{\theta_{23}(k_j)} }{\dot{m}_{33}(s(k_j)) m_{23}(s(k_j)) } \nu^2(y,t,k_j), \qquad n_1 < j \leq n_2, \; k_j \in E_1,
	\\ \nonumber
\underset{k_j}{\res} \nu^3(y,t,k) = &\; \frac{1}{\dot{m}_{33}(s(k_j))}\biggl(
\frac{m_{31}(S(k_j))e^{\theta_{13}(k_j)}}{W_1(k_j)}  \nu^1(y,t,k_j)
	\\  \label{nuresc}
&\hspace{0cm} + \frac{s_{11}(k_j)e^{\theta_{23}(k_j)} }{m_{23}(s(k_j))} \nu^2(y,t,k_j)\biggr), 
\qquad n_2 < j \leq n_3, \; k_j \in E_7,
	\\ \nonumber
\underset{k_j}{\res} \nu^2(y,t,k) = &\; \frac{m_{33}(s(k_j))m_{21}(S(k_j)) - m_{23}(s(k_j))m_{31}(S(k_j))}{\dot{W}_1(k_j) s_{11}(k_j)} e^{\theta_{12}(k_j)} \nu^1(y,t,k_j), 
	\\ \label{nuresd}
& \hspace{5cm} n_3 < j \leq n_4,\; k_j \in E_7,
	\\ \label{nurese}
\underset{k_j}{\res} \nu^3(y,t,k) = &\; \frac{m_{32}(S(k_j)) s_{11}(k_j) e^{\theta_{23}(k_j)} }{\dot{W}_2(k_j) m_{33}(s(k_j))} \nu^2(y,t,k_j), \qquad n_4 < j \leq N, \; k_j \in E_8,
\end{align}
\end{subequations}
}
where $\dot{f} := df/dk$ and
$$\theta_{ij}(k) = (l_i(k) - l_j(k))y + (z_i(k) - z_j(k))t.$$

\item For each zero $k_j$ in $E_1 \cup E_7 \cup E_8$, there are five additional points,
$$\omega k_j, \quad \omega^2 k_j, \quad \bar{k}_j, \quad \omega \bar{k}_j, \quad \omega^2 \bar{k}_j,$$
at which $\nu$ also has simple poles. The associated residues satisfy the residue conditions obtained from (\ref{nures}) via (\ref{nundef}) and the symmetries of Lemma \ref{Msymmlemma}. 
\end{itemize}
\end{theorem}
\proofbegin
The residue conditions (\ref{nures}) are obtained by multiplying the conditions in (\ref{Mres}) by $(1,1,1)$ from the left.
In order to derive (\ref{recoverq}) we note that (\ref{Phinatlambda0}) implies  
$$\nu(y,t,K_1) = \begin{pmatrix} 1 & 1& 1 \end{pmatrix} e^{\mathcal{L}(K_1)(x -y)} = \begin{pmatrix} e^{y-x} & 1 & e^{x-y} \end{pmatrix}.$$
Using (\ref{ydef}) and the relations
$$\frac{\partial}{\partial x} = q \frac{\partial}{\partial y}, \qquad
\frac{\partial}{\partial t}\bigg|_{\text{$y$ fixed}} = \frac{\partial}{\partial t}\bigg|_{\text{$x$ fixed}} + u \frac{\partial}{\partial x} ,$$
we find
\begin{align*}
& \frac{\partial}{\partial t}\log \bigl[\nu^1(y,t,K_1)\bigr]
= \biggl(\frac{\partial}{\partial t}\bigg|_{\text{$x$ fixed}} + u \frac{ \partial}{\partial x}\biggr)(y-x)
%= -q(x,t)u(x,t) + u (q -1) 
= -u(x,t),
%	\\
%& \frac{\partial}{\partial y}\log (\nu(y,t,K_1))_1 
%= \frac{1}{q} \frac{d}{dx} (y-x)
%= 1 - \frac{1}{q(x,t)},
	\\
& x = y - \log\bigl[\nu^1(y,t,K_1)\bigr].
\end{align*}
This gives the parametric representation (\ref{recoverq}).
\proofend

\appendix
\section{Proof of equation (\ref{qpositive})} \label{Aapp}
\renewcommand{\theequation}{A.\arabic{equation}}\nequation
We will show that the assumptions in (\ref{qassumptions}) imply that $u - u_{xx} + \kappa > 0$ for all $(x, t) \in \Omega$. 
Let $t^* > 0$. Let $\eta(x,t)$ be the unique solution of the differential equation
$$\eta_t(x,t) = u(\eta(x,t),t), \qquad  \eta(x, t^*) = x,$$
so that $t \mapsto (\eta(x,t), t)$ is the characteristic curve that passes through $(x,t^*)$ at time $t^*$.
The conservation law (\ref{conslaw}) implies that
\begin{align}\label{qeta}  
  \frac{d}{dt}[q(\eta(x,t), t) \eta_x(x,t)] = 0.
\end{align}
Moreover,
\begin{align}\label{etax}  
  \eta_x(x,t) = \exp\left(\int_{t^*}^t u_x(\eta(x, t'), t') dt'\right) > 0.
\end{align}  
Since $u \to 0$ as $x \to \infty$, every characteristic curve intersects either the initial half-line $\{x\geq 0, t = 0\}$ or the boundary $\{x= 0, 0 \leq t < T\}$. The assumptions (\ref{qassumptions}) imply that $q > 0$ at such an intersection point. 
Equations (\ref{qeta}) and (\ref{etax}) then imply that $q(\eta(x,t), t)$ is strictly positive on all of the characteristic curve. In particular, $q(x,t^*) > 0$. Since $t^* > 0$ was arbitrary, this proves (\ref{qpositive}).

\bigskip
\noindent
{\bf Acknowledgement} {\it The author acknowledges support from the EPSRC, UK.}

\bibliographystyle{plain}
\bibliography{is}

\end{document}